\documentclass[fleqn,usenatbib]{mnras}

\usepackage{newtxtext,newtxmath}

\usepackage[T1]{fontenc}
\usepackage{bm}
\usepackage{cleveref}
\usepackage{float}
\crefname{figure}{Fig.}{Figs.}
\crefname{table}{Table}{Tables}

\DeclareRobustCommand{\VAN}[3]{#2}
\let\VANthebibliography\thebibliography
\def\thebibliography{\DeclareRobustCommand{\VAN}[3]{##3}\VANthebibliography}

\usepackage{graphicx}	
\usepackage{amsmath}	
\usepackage{rotating}
\usepackage{xspace}
\usepackage{ulem}
\makeatletter 
  \patchcmd{\NAT@citex}
    {\@citea\NAT@hyper@{%
      \NAT@nmfmt{\NAT@nm}%
      \hyper@natlinkbreak{\NAT@aysep\NAT@spacechar}{\@citeb\@extra@b@citeb}%
      \NAT@date}}
    {\@citea\NAT@nmfmt{\NAT@nm}%
    \NAT@aysep\NAT@spacechar\NAT@hyper@{\NAT@date}}{}{}

  \patchcmd{\NAT@citex}
    {\@citea\NAT@hyper@{%
      \NAT@nmfmt{\NAT@nm}%
      \hyper@natlinkbreak{\NAT@spacechar\NAT@@open\if*#1*\else#1\NAT@spacechar\fi}%
        {\@citeb\@extra@b@citeb}%
      \NAT@date}}
    {\@citea\NAT@nmfmt{\NAT@nm}%
    \NAT@spacechar\NAT@@open\if*#1*\else#1\NAT@spacechar\fi\NAT@hyper@{\NAT@date}}
    {}{}
\makeatother

\newcommand{\Msun}{{\rm M_\odot}}
\newcommand{\msun}{{\,\rm M_\odot}}

\newcommand{\HI}{\ion{H}{I}\xspace}
\newcommand{\HeI}{\ion{He}{I}\xspace}
\newcommand{\HeII}{\ion{He}{II}\xspace}

\newcommand{\hii}{\ion{H}{II}\xspace}
\newcommand{\thesan}{\textsc{thesan}\xspace}

\newcommand{\thzoom}{\mbox{\textsc{thesan-zoom}}\xspace}
\newcommand{\thesandarkone}{\mbox{\textsc{thesan-dark-1}}\xspace}
\newcommand{\thesanone}{\mbox{\textsc{thesan-1}}\xspace}
\newcommand{\thesanHR}{\mbox{\textsc{thesan-hr}}\xspace}
\newcommand{\thesanVR}{\mbox{\textsc{thesan-variant}}\xspace}
\newcommand{\HM}{\ion{H}{$_2$}\xspace}
\newcommand{\HII}{\ion{H}{II}\xspace}
\newcommand{\HeIII}{\ion{He}{III}\xspace}

\title[Long-term imprints of reionization on galaxies]{The \thzoom project: Long-term imprints of external reionization on galaxy evolution}

\author[Zier et al.]{%
Oliver Zier,$^{1, 2}$\thanks{E-mail: \href{mailto:oliver.zier@cfa.harvard.edu}{oliver.zier@cfa.harvard.edu}}
Rahul Kannan,$^{3}$
Aaron Smith,$^{4}$
Ewald Puchwein,$^{5}$
Mark Vogelsberger,$^2$
Josh Borrow,$^{6}$
\newauthor
Enrico Garaldi,$^{7,8}$
Laura Keating,$^{9}$
William McClymont,$^{10, 11}$
Xuejian Shen,$^2$
and Lars Hernquist$^1$
\\
$^{1}$ Center for Astrophysics | Harvard \& Smithsonian, 60 Garden St, Cambridge, MA 02138, USA\\
$^{2}$ Department of Physics, Kavli Institute for Astrophysics and Space Research, Massachusetts Institute of Technology, Cambridge, MA 02139, USA\\
$^{3}$ Department of Physics and Astronomy, York University, 4700 Keele Street, Toronto, ON M3J 1P3, Canada \\
$^4$ Department of Physics, The University of Texas at Dallas, Richardson, TX 75080, USA \\
$^5$ Leibniz-Institut f\"ur Astrophysik Potsdam, An der Sternwarte 16, 14482 Potsdam, Germany \\
$^6$ Department of Physics and Astronomy, University of Pennsylvania, 209 South 33rd Street, Philadelphia, PA 19104, USA \\
$^7$ Kavli IPMU (WPI), UTIAS, The University of Tokyo, Kashiwa, Chiba 277-8583, Japan \\
$^8$ Institute for Fundamental Physics of the Universe, via Beirut 2, 34151 Trieste, Italy \\
$^9$ Institute for Astronomy, University of Edinburgh, Blackford Hill, Edinburgh, EH9 3HJ, UK \\
$^{10}$ Kavli Institute for Cosmology, University of Cambridge, Madingley Road, Cambridge CB3 0HA, UK\\
$^{11}$ Cavendish Laboratory, University of Cambridge, 19 JJ Thomson Avenue, Cambridge CB3 0HE, UK\\
}

\date{Accepted XXX. Received YYY; in original form ZZZ}

\pubyear{2025}

\begin{document}
\label{firstpage}
\pagerange{\pageref{firstpage}--\pageref{lastpage}}
\maketitle

\begin{abstract}
We investigate the impact of ionizing external ultraviolet (UV) radiation on low-mass haloes ($M_\text{halo}\lesssim 10^{10}\,\Msun$) at high redshift ($z\geq 3$) using $1140\,\Msun$ baryonic resolution zoom-in simulations of seven \thzoom regions.
We compare three simulation sets that differ in the treatment of external UV radiation: one employing a uniform UV background initiated at $z=10.6$ in addition to radiation transport for local sources, another with the background starting at $z=5.5$, and the default configuration using the large-scale radiation from the parent \thesanone simulation as a boundary condition.
The multi-phase interstellar medium (ISM) model, combined with its high mass resolution, allows us to resolve all star-forming haloes and capture the back-reaction of ionizing radiation on galaxy properties during the epoch of reionization.
When present, external UV radiation efficiently unbinds gas in haloes with masses below $10^9\msun$ and suppresses subsequent star formation. As a result, in simulations with early reionization, minihaloes fail to form stars from pristine gas, leading to reduced metal enrichment of gas later accreted by more massive haloes. 
Consequently, haloes with masses below $10^{10}\msun$ at all simulated epochs ($z>3$) exhibit lower metallicities and altered metallicity distributions.
The more accurate and realistic shielding from external UV radiation, achieved through self-consistent radiative transfer, permits the existence of a cold but low-density gas phase down to $z=3$.
These findings highlight the importance of capturing a patchy reionization in high-resolution simulations targeting high-redshift galaxy formation. We conclude that at minimum, semi-numerical models that incorporate spatially inhomogeneous reionization and a non-uniform metallicity floor are necessary to accurately emulate metal enrichment in minihaloes.
\end{abstract}




\section{Introduction}
Around 380,000 years after the Big Bang (redshift $z\approx 1100$), the Universe had expanded enough to cool to around 3000\,K. This allowed protons and electrons to recombine, leading to the decoupling of baryonic matter from the radiation field, which is known as the Cosmic Microwave Background \citep[CMB;][]{alpher1948relative,penzias1979measurement}.
After this epoch of recombination, the Universe entered an era known as the cosmic dark ages, characterized by a lack of significant sources of radiation and ionization.
During this era, small inhomogeneities in the density field were able to grow due to the gravitational force \citep{Smoot1992}, and eventually formed the first stars around redshift $z \approx 30$ \citep{Klessen2023}. 
These early Population-III stars formed from metal-free gas and emitted large amounts of Lyman continuum photons (LyC; $h\nu \geq 13.6\,\mathrm{eV}$), capable of ionizing hydrogen in their surrounding areas.
As matter continued to collapse, the first galaxies began to form \citep{Bromm2011}. Reaching higher levels of star formation, these galaxies were responsible for creating expanding ionized bubbles in the intergalactic medium (IGM).  These bubbles overlapped in a spatially inhomogeneous process termed ``patchy reionization'' \citep{shapiro1986cosmological,haardt1995radiative,gnedin1997reionization,madau1999radiative,gnedin2000effect}.
Although early studies placed the completion of reionization around or above $z= 6$ \citep{fan2006constraining}, more recent observations indicate it may have ended at somewhat lower redshifts \citep{becker2015evidence,Kulkarni2019, bosman2022hydrogen}.

Reionization has had a significant impact on both the IGM and the properties of galaxies, especially for lower-mass haloes. A common distinction drawn for regions in the early universe is between those that reionize internally and those that reionize externally. Internally reionized regions generate the majority of their ionizing photons, while externally reionized regions are primarily ionized by nearby sources. 
Smaller haloes with masses $\sim 10^9 \msun$ or lower, are particularly affected by photo-heating, which heats the gas to temperatures of around $10^4$\,K, which is of the order of their virial temperatures.
As a result, ionized gas can escape the potential wells of these haloes, and gas inflow can be suppressed \citep[e.g.][]{Rees1986, Shapiro2004, Okamoto2008}.
External reionization introduces an additional environmental factor in galaxy formation, potentially inducing permanent or temporary quenching of these galaxies \citep{Okamoto2008, Wu2019a, gutcke2022lyra}.
The time of quenching is imprinted in the stellar population of these small haloes as an age cutoff.

The advent of the James Webb Space Telescope (JWST) has offered a detailed observational view of the evolution of galaxies during the Epoch of Reionization (EoR).
Its initial results are partially in tension with theoretical expectations from standard galaxy-formation models \citep{BoylanKolchin2023, Kannan2023}, as, e.g. more massive and bright galaxies at $z>8$ \citep{Labbe2023, Finkelstein2023, Weibel2024} are observed. This tension could be, on the one hand, solved by modifications of the baryonic model such as less efficient feedback at high redshift \citep[e.g.][]{Dekel2023, Yung2024} or bursty star formation \citep{Shen2023, Sun2023, Semenov24a, Semenov24b}, or on the other hand by more exotic modifications of the underlying physics model, such as early dark energy \citep{EDE2022, Shen2024EDE}, primordial black holes \citep{Liu2022} or cosmic string loops \citep{Jiao2023, Koehler2024}. 
To accurately distinguish between the different proposed solutions,
it is important to model the relevant physical processes as accurately as possible, including exploring radiative feedback at high resolution. 

Although it is possible to study reionization by performing radiative transfer (RT) in post-processing \citep{ciardi2003simulating, iliev2007kinetic, mcquinn2007morphology, mcquinn2009he}, only fully-coupled radiation hydrodynamics (RHD) simulations that model individual galaxies can self-consistently account for feedback from the radiation on the baryonic matter.
RHD simulations are significantly more computationally expensive than pure hydrodynamic ones because the speed of light sets the maximum signal speed. 
Due to the additional constraint to resolve individual small-scale star-forming regions within galaxies as sources of ionizing radiation, simulations have to balance between the simulation volume, resolution, and physics modelling \citep{CROC2014,codai,codaii,codaiii, SPHINX2018, SPHINX2022, Obelisk2021, Aurora2017, Finlator2018,xu2016galaxy,wells2022, bhagwat2023spice, Bhagwat2024}.
The \thesan project \citep{Thesan1, ThesanAaron, ThesanEnrico, ThesanDR}, for instance, achieved a relatively large volume of side length $95.5 \, \text{cMpc}$ by combining the IllustrisTNG galaxy formation model \citep{Weinberger2017, Pillepich2018Model} with a novel radiative transfer module \citep{arepoRT} in the moving-mesh code {\small AREPO} \citep{springel2010pur, weinberger2020arepo}.
IllustrisTNG successfully reproduces galaxies at low redshift \citep{Springel2018, Marinacci2018, Naiman2018, Nelson2018, pillepich2018first, TNGPublicDataRelease} and is based on the previous Illustris galaxy formation model \citep{IllustrisModel2013, IllustrisNature, IllustrisIntro,genelIllustris}.
Although \thesan captured seven orders of magnitude in spatial scale, it only resolved haloes with masses above $10^9\msun$ with more than 300 particles, limiting its ability to explore the evolution of minihaloes.

Cosmological simulations that do not focus on the EoR itself but on lower redshift typically use a uniform ultraviolet background \citep[UVB, e.g.][]{FG09,Haardt2012,Puchwein2019, faucher2020cosmic} along with approximate self-shielding prescriptions, which are often based solely on the local gas density \citep{Rahmati2013}. 
Examples of large-volume simulations using this setup include Illustris \citep{IllustrisNature, IllustrisIntro}, IllustrisTNG \citep{pillepich2018first,nelson2019first}, MillenniumTNG \citep{MTNG_main}, EAGLE \citep{schaye2015eagle}, {\small FLAMINGO} \citep{schaye2023flamingo} and Magneticum Pathfinder \citep{dolag2016sz}. 
Additionally, it is possible to use semi-analytic models to emulate a patchy reionization, as demonstrated in the {\small ASTRID} simulation \citep{bird2022astrid,ni2022astrid}.
\cite{ThesanEnrico} compared the fiducial \thesan model with on-the-fly RT with an equivalent run using a uniform UV background. At the end of the EoR, they found only minor differences for haloes with stellar masses $> 10^6 ~{\rm M}_\odot$, for which internal sources dominate the local radiation field.

The \thesanHR \citep{borrow2023thesan, Shen2024} project evolved significantly smaller boxes with side lengths of $5.9 \, \text{cMpc}$ and $11.8 \, \text{cMpc}$, but at up to 64 times higher mass resolution than the main \thesan run.
It used the same physics model as \thesan, though the smaller box size modified the global reionization history in the RHD run.
The simulations resolved haloes with total masses as low as $10^7 \msun$.
Significantly different results were found for these small haloes when comparing the RHD simulation with the same simulation setup but with a uniform UVB.
In the latter scenario, star formation in haloes below $10^9 \msun$ was almost immediately shut off after the initialization of the UVB.
This resulted in a reduced stellar-to-halo mass ratio and baryon-to-halo mass ratio during the end of EoR in the UVB run and the end of the EoR in the RHD simulation, along with larger galaxy sizes and different stellar age distributions.
Notably, the RHD simulation predicted the formation of metal-free stars at redshifts as low as $z=5$ in small haloes, which were prevented from forming stars in the uniform UVB run.
These results are important for higher resolution simulations capable of resolving such small haloes, such as {\small FIREBOX} \citep{firebox} or TNG-50, but do not use RHD \citep{nelson2019first}.

\thesan and \thesanHR both used the effective equation of state model from \cite{springel2003cosmological}, which assumes an equilibrium in the interstellar medium (ISM) between cooling and heating by stellar feedback.
Although this model can successfully reproduce galaxy properties at low redshifts by using observational constraints to tune several free parameters, it cannot resolve the multiphase structure of the ISM and, therefore, the small-scale interaction of photons with it. 
An alternative model implemented in the {\small AREPO} code is the {\small SMUGGLE} model \citep{Marinacci2019}. Similar to the {\small FIRE} model \citep{Hopkins2018}, {\small SMUGGLE} aims to resolve all three phases of the ISM.
By coupling RT with this model, effects such as self-shielding can be modelled more accurately \citep{Kannan2020b}.

This paper is part of a series introducing the \thzoom simulation project, which consists of high-resolution zoom-in simulations of haloes from the original \thesan volume.
It uses the {\small SMUGGLE} model with seven radiation bins and extends the original chemical network from \thesan by incorporating molecular hydrogen and dust chemistry.
All haloes are evolved to redshift $z=3$ and, by default, use the large-scale radiation field of the parent box as the boundary condition for the high-resolution regions.
This approach provides a consistent treatment of external reionization, although we note that the external radiation field is based on the parent simulation which used the effective equation of state ISM model.
An overview of the simulation project is provided in \cite{zoomIntro}. It was used to study the star formation efficiency \citep{Shen2025, Wang2025}, the star-forming main sequence \citep{McClymont2025a}, galaxy sizes \citep{McClymont2025b} and Pop III star formation \citep{ZierPopIII} in the high redshift universe. 
Besides the simulations with the radiation background from the parent simulation, additional simulations use a uniform (except for self-shielding) UVB starting at the redshift $z=10.6$ and one starting at redshift $z=5.5$, the end of the EoR in \thesanone.
The additional simulation variants should emulate early and late external reionization scenarios.

This paper aims to compare the predictions of these modified runs of the same haloes and analyze the influence of patchy reionization compared to uniform reionization on galaxy properties.
The higher resolution, extension to redshift $z=3$, and the ability to resolve the multiphase ISM structure make the \thzoom suite ideal for this task. 
This paper is structured as follows:
In Section~\ref{sec:methods}, we introduce the \thzoom project focusing on the different treatments of the background field.
In Section~\ref{sec:UVBGalaxyEvolution}, we compare the general evolution of different global galaxy properties, such as baryon and stellar-to-halo mass evolution, in the different simulation sets.
In Section~\ref{sec:UVBGasPhase}, we have a closer look at the influence of the radiation field on the gas structure within galaxies, and in Section~\ref{sec:UVBStellarEvolution}, we compare the structure of the stellar component in the different sets. 
We discuss and summarize our results in Section~\ref{sec:summary}.

\section{Methods}
\label{sec:methods}
The \thzoom project, including a detailed description of the numerical details, is introduced in \cite{zoomIntro}, to which we also refer for more detailed information.
In \cref{subsec:overThZoom}, we will provide an overview of the complete simulation suite. Additionally, in \cref{subsec:BackgroundField}, we will discuss the standard treatment of the large-scale radiation field and its modifications.

\subsection{Overview of the \thzoom project}
\label{subsec:overThZoom}
The \thzoom suite includes zoom-in simulations of 14 different dark matter (DM) haloes, which were drawn from the \thesandarkone simulation, which is the dark matter counterpart to the flagship \thesanone simulation from the \thesan project \citep{Thesan1, ThesanAaron, ThesanEnrico}.
The initial conditions for the zoom-in simulations were created for haloes selected at $z=3$ to encompass a halo mass range from $10^8 \msun$ to $10^{13}\msun$. The high-resolution regions cover a sphere with a radius of about four times the virial radius of the central halo at $z=3$.
 This condition implies that, at high redshift, the high-resolution regions cover a considerably larger volume than just the immediate surroundings of the central haloes, which we will use later to enhance our statistical analysis.

 All simulations use the massively parallel {\small AREPO} code
\citep{springel2010pur,pakmor2016improving,weinberger2020arepo} which employs a moving, unstructured, Voronoi mesh together with a second-order accurate finite volume method to solve the Euler equations.
The cells within the mesh move approximately with the local fluid velocities, leading to approximately constant mass resolution in the zoom-in region.
The code allows for the splitting/merging of cells whose mass deviates by more than a factor of 2/0.5 from a predefined target mass $\rm m_{gas}$ to ensure the later condition.
To compute gravitational forces, the codes use the hybrid TreePM method \citep{Bagla2022}, which relies on a hierarchical octree \citep{Barnes1986} for calculating short-range forces, while the Fourier-based particle-mesh method is used for long-range forces  \citep{Aarseth2003}.
The extension {\small AREPO-RT} \citep{arepoRT} further solves hyperbolic conservation laws on the Voronoi mesh for the zeroth and first moments of radiation intensity, specifically the photon number density and photon flux, together with the M1 closure relation \citep{Levermore1984, Dubroca1999}.
To enhance the code's scalability and reduce the overhead of interprocess communication, the ``Node-to-Node'' communication scheme from \cite{zier2024adapting} is used.
The radiation field is discretized into seven frequency bins, which represent the following spectral ranges: infrared (IR, $0.1-1$ eV), optical ($1.0-5.8$ eV), far-UV ($5.8-11.2$ eV), Lyman-Werner (LW, $11.2-13.6$ eV), the hydrogen ionizing band ($13.6-24.6$ eV), and two helium ionizing bands ($24.6-54.4$ eV, $>54.4$ eV). 

The radiation field interacts with the baryonic matter within a non-equilibrium chemical network \citep{Kannan2020b}, which evolves the abundances of the primordial species $\HM, \HI, \HII, \HeI, \HeII,$ and $\HeIII$. 
The heating and cooling rates from these primordial species are calculated self-consistently. For metal-line cooling, we utilize tables that assume ionization equilibrium with the UV background from \cite{FG09}, scaling them linearly with the gas metallicity \citep{IllustrisModel2013}.
The code additionally accounts for photoelectric heating from Far-UV photons \citep[$5.8-11.2 \,  \mathrm{eV}$, based on][]{Wolfire2003}, cooling due to gas-dust interactions \citep[based on][]{Burke1983} and Compton cooling/heating off the CMB.
Following \cite{McKinnon2016, McKinnon2017},
cosmic dust is represented as an additional scalar property of the gas cells, which neglects the relative velocities between gas and dust.
This dust is produced by supernovae and winds from asymptotic giant branch (AGB) stars, grows in the dense ISM and is destroyed by supernova shocks and sputtering.
The scalar dust modifies the cooling and heating rates by interacting with the infrared radiation bin  (IR, $0.1-1$ eV) \citep{Kannan2021}.

The star formation and stellar feedback routines for supernovae and stellar winds are modified versions of the original SMUGGLE model \citep{Marinacci2019}.
Gas cells smaller than their thermal Jeans length and denser than $n_\mathrm{H} = 10~\mathrm{cm}^{-3}$ are allowed to form stars.
Star formation is implemented probabilistically, with a relatively high efficiency of 100\%  per free-fall time, which prevents unresolved gas from collapsing to artificially high densities \citep{Hu2023}.
Stellar particles represent a single stellar population that follows the Chabrier Initial Mass Function \citep[IMF, ][]{chabrier2003galactic} with minimum and maximum masses of $0.1~\mathrm{M}_\odot$ and $100~\mathrm{M}_\odot$ respectively.
The IMF is used to stochastically determine the timing of supernova events for all stars above $8\msun$.
To ensure that typically only one supernova occurs per integration time step for each star particle, a specific time step criterion is employed. Each supernova event injects the canonical energy of $10^{51}~\mathrm{ergs}$ into the surrounding ISM.
Due to insufficient resolution to accurately capture the Sedov-Taylor phase, direct energy injection could result in an overcooling problem \citep{DV2012}. 
Therefore, we instead inject the terminal momentum at the transition to the momentum-conserving phase \citep{Marinacci2019}.
We use analytic prescriptions from \cite{Hopkins2018}, based on the \textsc{Starburst99}~\citep{Leitherer1999} stellar evolution model, to calculate the mass loss rate and energy from stellar winds of young OB and asymptotic giant branch (AGB) stars.
The metal enrichment rates are taken from \cite{IllustrisModel2013}, and we set an initial metal mass fraction of  $Z = 10^{-8}$.
We use an additional feedback channel that injects momentum during the first $5~\mathrm{Myrs}$ before the first supernova events occur. This approach helps to match better the stellar-mass-halo-mass relations at high redshift \citep[e.g.][]{Moster2018, Behroozi2019}, as further motivated in \cite{zoomIntro}. 

Each stellar particle continuously injects photons into the 16 nearest Voronoi cells.
The age and metal-dependent luminosities are calculated using the Binary Population and Spectral Synthesis models (BPASS v2.2.1; \citealt{BPASS2017}). These models are also employed to compute averaged quantities, such as the ionization cross sections per frequency bin. For this calculation, we assume a $2$\,Myr old spectrum at quarter solar metallicity.
Additionally, we apply the corrections described in \cite{deng2024} to reduce the effect of underresolved  \hii regions.

In addition to the simulations using the default physics implementation presented in this section, the \thzoom suite also includes simulations with modified physical parameters. 
Examples include simulations with lower star formation efficiencies and those conducted without early stellar feedback.
For a comprehensive overview, we refer to \cite{zoomIntro}; however, in this paper, we focus solely on the standard simulations and those that use a modified treatment of the large-scale radiation field, as discussed in the following section. 
These specifically modified simulations exist only at one resolution, the zoom factor 8 (see \cref{table:res} for details on mass resolution), and only for the seven smallest haloes that are expected to be most affected by reionization.
We describe the physical properties of these seven zoom-in regions in \cref{table:overviewSimulations}.
In the main part of the paper, we exclusively use the zoom factor 8 simulations. 
The simulations based on the default \thzoom model also exist for zoom factor 4, which we use in Appendix~\ref{app:resolution} to analyze the impact of numerical resolution on our results.

\begin{table}
	\centering

	\begin{tabular}{lcccccccc} 
		\hline
		Zoom & $m_\mathrm{DM}$ & $m_\mathrm{gas}$ & $\epsilon_\mathrm{DM, stars}$ & $\epsilon_\mathrm{gas}^\mathrm{min}$\\  
		factor&  [$\mathrm{M}_\odot$] & [$\mathrm{M}_\odot$] & [cpc] & [cpc]\\
		\hline
		8x  & $6.09 \times 10^3$ & $1.14 \times 10^3$ & $276.79$ & $34.60$ &\\
		4x &  $4.86 \times 10^4$ & $9.09 \times 10^3$ & $553.59$ & $69.20$\\
		\hline
	\end{tabular}
 \caption{Numerical resolution for the simulations from the \thzoom project analyzed in this paper. 
     The columns, from left to right, indicate the spatial zoom factor in relation to the parent box, the mass of the high-resolution dark matter particles and gas cells, and the (minimum) softening length for both gas cells and dark matter particles. 
     In the main body of this paper, we focus on simulations with a zoom factor of 8; however, the 4x simulations are used in Appendix~\ref{app:resolution} for a convergence study.}
	\label{table:res}
\end{table}

\subsection{Consistent treatment of the large scale radiation field}
\label{subsec:BackgroundField}
Smaller simulation boxes experience delayed reionization due to the absence of massive haloes and the impossibility of external reionization \citep[e.g.][]{borrow2023thesan}. 
To address this issue, the \thzoom project uses the radiation field from the \thesanone parent RHD simulation as the boundary condition for the high-resolution region in its default simulation set, referred to as ``standard''.
In practice, in each time integration step, the photon density and photon flux of active low-resolution Voronoi cells are set to precalculated values. 
These values are determined through linear interpolation in time between the high-cadence outputs from the parent box prior to redshift $z=5.5$. 
Between $z=5.5$ and $z=5$, there is a smooth transition from the last output from \thesanone to the uniform UV background from \cite{FG09}, which we also use at lower redshift.
In this paper, we analyse two additional simulation sets that do not modify the radiation field in the low-resolution region. Instead, they add in the chemical network the photons of the uniform UV background from \cite{FG09} along with the self-shielding approximation from  \cite{Rahmati2013}:
In the first simulation set (``UVB''), this background is applied in addition to the local radiation field starting from redshift $ z=10.6$, while in the second one (``lateUVB''), it is only applied after redshift $z=5.5$.
We note that the self-shielding is only applied in both cases for $z< 6$, consistent with previous simulations such as Illustris and its successors.
After redshift $z=5.5$, the radiation field treatment in both additional sets is similar to the ``standard'' run, except that the background is directly applied to all gas cells. In the ``standard'' case, photons must first diffuse into the high-resolution region.
As we will demonstrate later, the second approach allows for more accurate modelling of self-shielding.
It is important to note that a radiation background is only used for the three hydrogen and helium ionising radiation bins, which are the only bins included in the \thesanone simulation. 
Only local sources in the high-resolution regions are considered for the remaining radiation bins. 
We compare in  \cref{app:UV_strength} the strength of the median UV background for each simulation set.
\begin{table}
	\centering
	
	\begin{tabular}{l|cccc} 
  \hline

  Name & $M_{\rm halo} \left(z = 3\right)$  & $\rm V_{hr} (z = 3)$ & $\rm V_{hr} (z = 8)$ &$z_{\rm reion}$\\
  & $\left[\msun \right]$ &$\rm  \left[cMpc^3\right]$ &$\rm  \left[cMpc^3\right]$\\
  \hline
  m$10.4$ & $2.53 \times 10^{10} \msun$& 1.088& 2.185 & 7.0\\ 
  m$10.0$ & $1.07 \times 10^{10} \msun$ &0.479 &1.553 & 7.1\\ 
  m$9.7$ &$4.58 \times 10^9 \msun$ & 0.199& 0.770 & 7.7\\ 
  m$9.3$ &$1.95 \times 10^9 \msun$ & 0.129 & 0.312 & 6.6\\ 
  m$8.9$ &$8.29 \times 10^8 \msun$ & 0.075 & 0.204 & 6.6\\ 
  m$8.5$ & $3.51 \times 10^8 \msun$& 0.069& 0.272  & 8.0\\
  m$8.2$ & $1.52 \times 10^8 \msun$& 0.060& 0.066  & 6.8\\
\hline

 \end{tabular}%
    \caption{The properties of the target haloes of each zoom-in region that are discussed in this paper.
    We analyze three simulations for each halo that differ in their treatment of the radiation boundary condition (see \cref{subsec:BackgroundField}). The zoom factor for these simulations is 8x, as detailed in Table \ref{table:res}. 
    We show the name of the target halo, its mass at $z=3$ in the \thesandarkone simulation, the comoving volume of high-resolution cells at $z=3$ and $z=8$ in the simulation with ``standard'' boundary conditions, and the critical redshift $z_{\rm reion}$ defined as the time when the volume-weighted mean neutral hydrogen fraction of the full zoom-in region in the ``standard'' simulation set falls below $0.1$. For most analyses, we consider not only the target haloes but all haloes composed of only high-resolution particles. }
    \label{table:overviewSimulations}
\end{table}

\begin{figure*}
    \centering
    \includegraphics[width=1\linewidth]{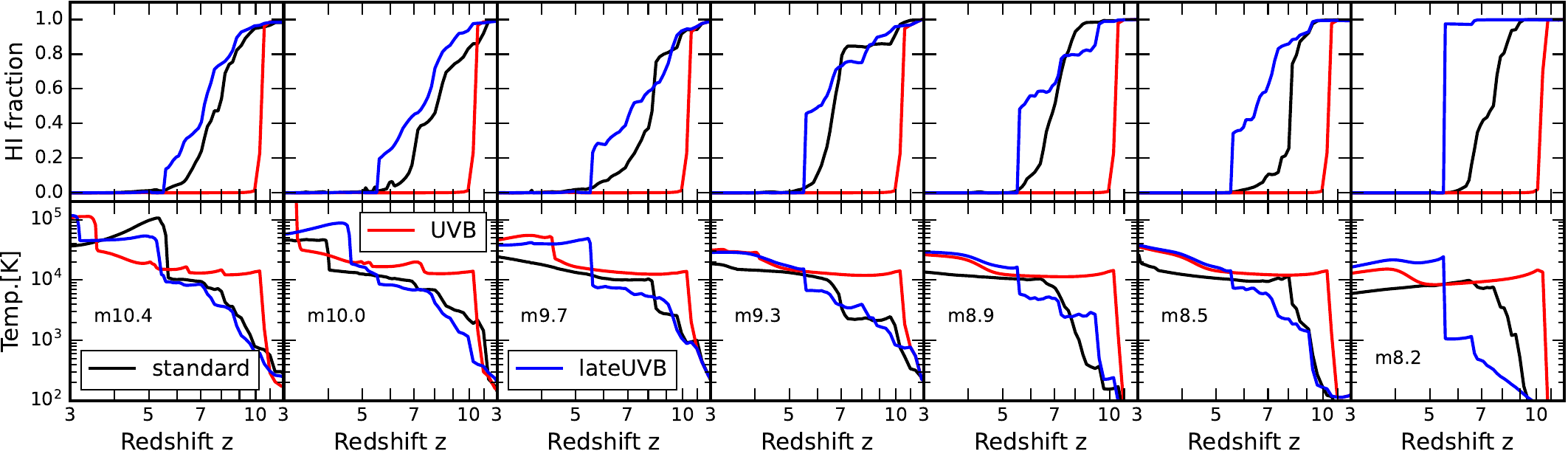}
    \caption{The reionization history of the high-resolution IGM (all gas not associated with a halo) for all seven zoom-in regions. We present the mass-weighted mean neutral hydrogen fraction (top) and temperature (bottom). 
    In the simulation with a uniform ultraviolet background (UVB) (shown in red), reionization occurs almost instantaneously at redshift $z=10.6$. In contrast, we find a gradual ionisation of the IGM from local sources for the other two cases.
    Notably, for small haloes, we can observe the onset of outside-in reionization in the standard case (depicted in black) as seen by comparing to ``lateUVB''. Generally, the simulations with a delayed UVB (shown in blue) experience reionization at a later stage. However, with the activation of the uniform UVB at $z=5.5$,  the IGM becomes fully ionised.
   }
    \label{fig:reionizaiton_history}
\end{figure*}

\subsection{Expected impact from external radiation}
The external UV radiation can ionize the gas in haloes, heating it to a canonical temperature of  $\approx 10^4$\,K.
In low-mass haloes, this temperature can exceed the virial temperature given by \citep[e.g.][]{Mo2010}:
\begin{equation}
    T_{\rm vir} = \frac{1}{2} \frac{\mu m_{\rm H}}{k_{\rm B}} \frac{G M_{\rm vir}}{R_{\rm vir}},
    \label{eq:virialTemperature}
\end{equation}
where $\mu$ is the molecular weight per particle ($\mu \approx 0.6$ for fully ionized gas), $m_H$ the mass of one hydrogen atom, $k_B$ the Boltzmann constant, $G$ is Newton's gravitational constant, and $M_{\rm vir}$ and $R_{\rm vir}$ are the virial mass and radius of the halo, respectively. 
It is important to note that this equation is an approximation; for instance, a significant contribution from non-thermal kinetic energy may require modifications \citep{Lochhaas2021}.
Gas that is heated above the virial temperature can escape the halo, which can significantly reduce star formation. Even in higher-mass haloes that are not entirely quenched by ionizing radiation, the inflow rates from the circumgalactic medium (CGM) to the ISM can still be reduced, even if the gas remains bound to the halo ~\citep[e.g.][]{Rees1986, Shapiro2004, Okamoto2008}.
The virial temperature is not only a function of the halo mass but also of the halo radius. Nevertheless, as we will show later, a virial temperature of $\rm O(10^4K)$ corresponds to a halo mass in the range of  $10^8 \msun$ to $10^9\msun$.

As previously discussed, the redshift for the onset of external radiation varies depending on the type of simulation set.
For simulations with uniform UV background, the critical redshift is $z=10.6$ (for ``UVB'' runs) and $z=5.5$ (for ``lateUVB'' runs).
For the ``standard'' set with the radiation background from the \thesanone box, the critical redshift depends on the position of the zoom-in region.

We present the reionization history of the high-resolution intergalactic medium (IGM) in \cref{fig:reionizaiton_history}. The IGM is defined as all gas that is not associated with any halo. 
In the ``UVB'' simulations, we observe a rapid ionization and temperature increase to around $10^4$\,K at $z=10.6$.
In simulations featuring a more massive central halo, we see additional heating caused by \HeII reionization and stellar feedback.
In most of the ``lateUVB'' simulations, which have a delayed external radiation field, we note a gradual process of internal reionization that rapidly completes at $z=5.5$ when the external radiation field is activated.
For the halo m8.2, the internal sources are insufficient to ionize a significant portion of the IGM, though there is some gravitational heating during structure formation and heating from stellar feedback.
For the ``standard'' simulations that incorporate the radiation background from the \thesanone simulation, reionization is driven by a combination of internal and external sources. The presence of external sources speeds up the reionization process compared to the ``lateUVB'' runs and generally leads to a smoother transition. By the end of the parent simulation at $z=5.5$, all volumes are fully ionized.

\subsection{Analysis method}
\label{subsec:anaylsisMethod}

For each simulation, 189 snapshot files were generated, covering redshifts from $z=16$ to $z=3$, with a time cadence of around $10~\mathrm{Myrs}$. These snapshots contain the most important quantities for all gas, dark matter and stellar particles.
Using the friends-of-friends (FOF) algorithm \citep{Davis1985}, we created a halo catalogue for each snapshot, which was further processed with the SUBFIND-HBT \citep{springel2001populating, Gadget4} algorithm.
The FOF algorithm considers only the spatial proximity of particles, whereas the SUBFIND-HBT algorithm identifies gravitationally bound structures by accounting for the gravitational, relative kinetic, and thermal energy of the particles.
This halo catalogue allows us to trace the target zoom-in halo back in time, starting at $z=3$. It is important to note that multiple progenitors may exist, particularly in cases of major mergers. 

The criterion that particles within four times the virial radius of the target halo at $z=3$ are high-resolution particles results in an extended high-resolution region at higher redshifts.
This region includes additional, typically smaller haloes consisting solely of high-resolution particles. We can incorporate these smaller haloes into our analysis to enhance the statistical reliability of our results. Consequently, we also provide the size of the high-resolution Lagrangian region at both low and high redshifts in  \cref{table:overviewSimulations}.
For some figures, we combine the data from several snapshots to increase the statistical sample further, especially at the higher mass end. 
In this case, we will explicitly state the redshift range used.

\section{Influence of the UV background on averaged galaxy properties}
\label{sec:UVBGalaxyEvolution}
\begin{figure*}
    \centering
    \includegraphics[width=0.95\linewidth]{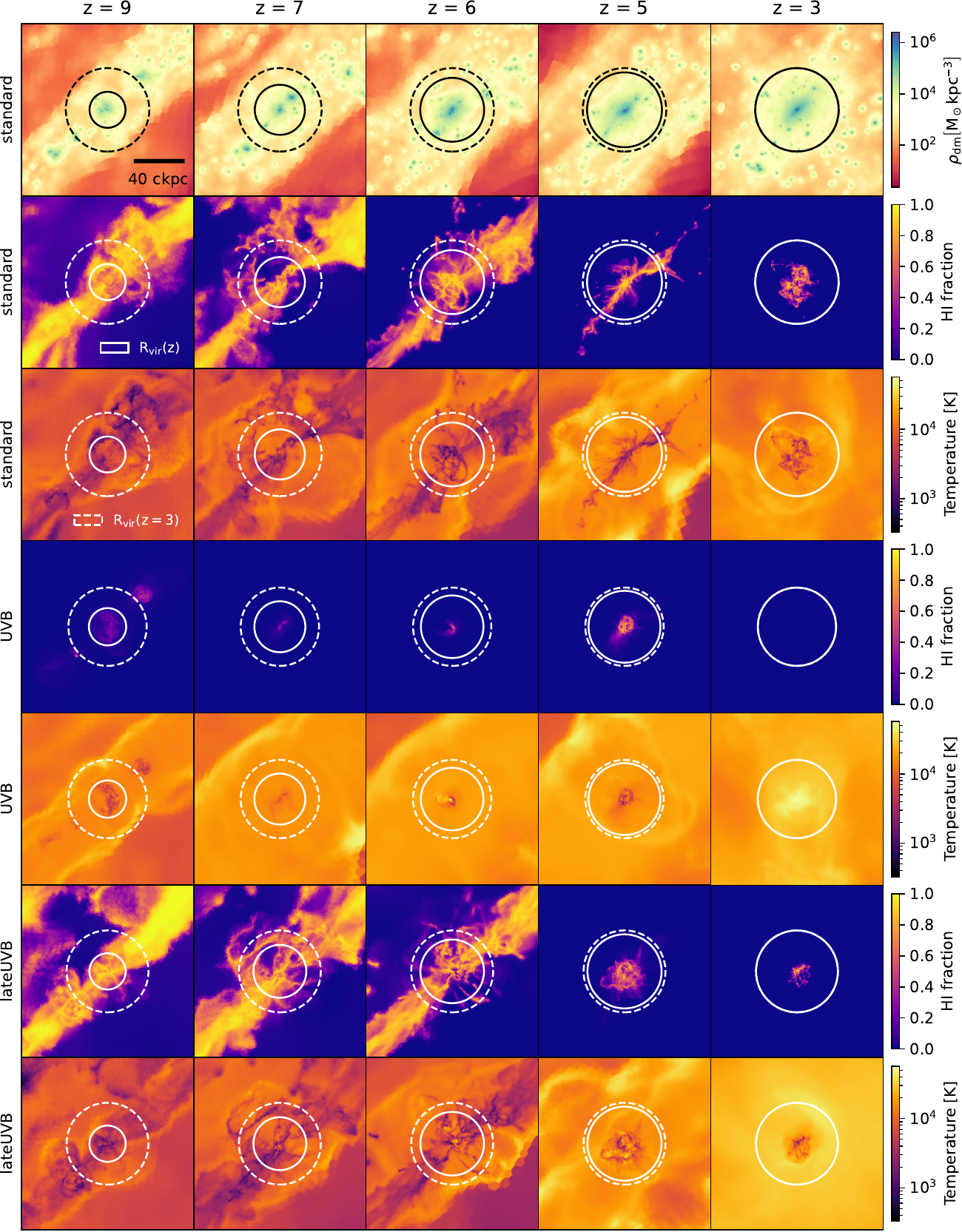}
    \caption{The dark matter density (row 1), neutral hydrogen fraction (rows 2, 4, 6), and gas temperature (rows 3, 5, 7) around the halo m9.3 are shown for five different redshifts.
    The gas properties are shown for all three implementations of the radiation background. The dark matter density is displayed only for the ``standard'' simulation, as it is minimally affected by the radiation field. 
    All quantities are mass-weighted means taken from a slice in the z-direction, measuring $\rm 75\,ckpc$ in size. The comoving virial radius is shown for the redshift of the plot (solid line) and at $z=3$ (dashed line) of the central halo. 
    The halo is located within a cosmic filament that is almost instantaneously ionized at the onset of the uniform UV background in the ``UVB'' and ``lateUVB'' simulations.
In contrast, with the more realistic boundary conditions provided by the \thesanone simulation, ionization occurs gradually.}
    \label{fig:goodPic}
\end{figure*}
\begin{figure*}
    \centering
    \includegraphics[width=1\linewidth]{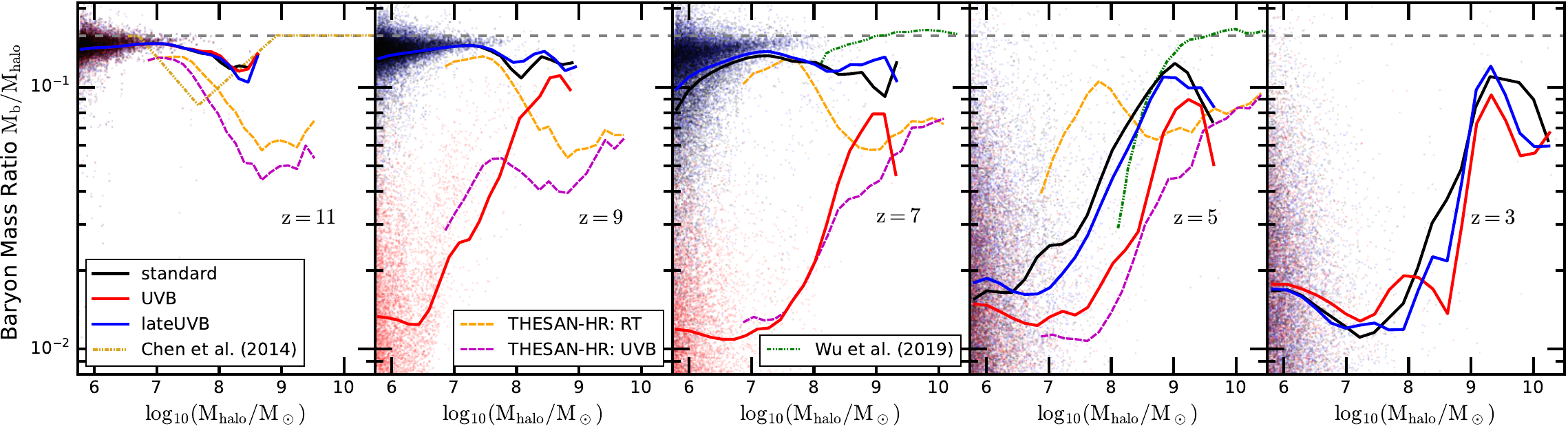}
    \caption{The evolution of the baryon mass to halo mass ratio over cosmic time is analyzed for three sets of simulations, each differing in their treatment of the background radiation field. 
    Each set consists of seven simulations, and we show the results for all haloes (individual points) as well as the mean value within the binned halo mass (solid lines). The dashed line represents the cosmic baryon fraction of $0.157$. For comparison, we include additional values from the highest resolution of the \thesanHR suite, specifically the L4N512 simulation \citep{borrow2023thesan}. 
    This suite consists of one simulation with radiative transfer (orange line, RT) and another with a uniform UVB (magenta line, UVB). 
    Furthermore, we provide results from \protect\cite{Chen2014} (cyan line) at redshift $z=11$ and from \protect\cite{Wu2019a} (green lines) at $z=5$ and $z=7$.
    In the ``UVB'' set, the UVB results in a loss of baryons for halo masses below $10^9\msun$. However, after reionization, our models predict a similar baryon content across all simulations.
    For better statistics we use the five snapshots closest to the target redshift, resulting in a time range of around $40$\,Myr.
    }
    \label{fig:baryonFraction}
\end{figure*}

\begin{figure*}
    \centering
    \includegraphics[width=1\linewidth]{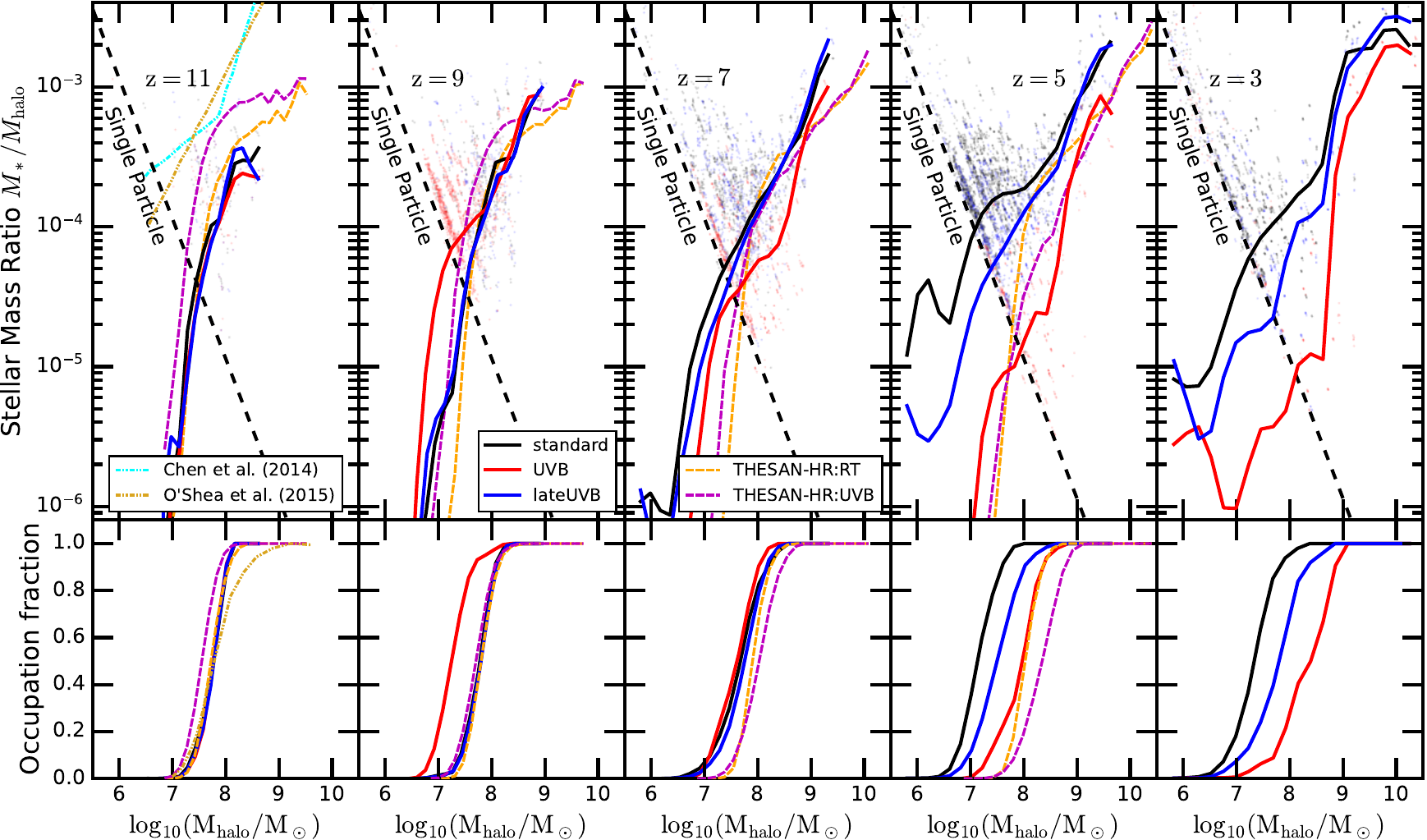}
    \caption{The evolution of the stellar mass to halo mass relation through cosmic time for three sets of simulations, which differ only in their treatment of the external radiation field. Each set consists of seven simulations, and we present the results for individual haloes (dots) as well as the mean in bins of the halo mass (solid line). The black dashed line illustrates the expected relationship for haloes that consist of only one stellar particle, with a typical mass of $1.14\times 10^3\msun$ for a zoom factor 8.
    The lower row shows the fraction of haloes containing at least one single star particle as a function of halo mass.
    We discuss the impact of resolution and different definitions of the occupation fraction based on the stellar mass threshold in \cref{app:resolution}.
    At redshift $z=3$, all models show similar behaviour for haloes with masses above $5\times 10^9\msun$.
However, for lower halo masses, the occupation fraction varies depending on how the radiation background is treated. For comparison, we also present results from the \thesanHR simulation suite \citep{borrow2023thesan} which we recalculated using the original simulation output and compare our results with those from \protect\cite{Chen2014} and \protect\cite{shea2015}.
    For better statistics we use the five snapshots closest to the target redshift, resulting in a time range of around $40$\,Myr.}
    \label{fig:stellarFraction}
\end{figure*}
To gain a first visual impression of the impact of the external radiation on haloes and their surroundings, we show the mass-weighted mean dark matter density, neutral hydrogen fraction and gas temperature in a slice of depth $\rm 75\,ckpc$ at different redshifts in \cref{fig:goodPic} for the halo m9.3. 
As indicated by the reionization history in \cref{fig:reionizaiton_history}, the high-resolution region in this simulation is partially ionized by internal sources, as the neutral hydrogen fraction in the ``lateUVB'' simulation has already dropped to around 50\% before the onset of the external radiation at $z=5.5$.
The central halo lies within a dark matter filament that contains many smaller haloes and shows an increased gas density.
The filament becomes fully ionized at $z=10.6$ upon the onset of the uniform UVB in the ``UVB'' simulation, which also heats the gas to $>10^4$\,K. Up until $z=5$, a small amount of neutral hydrogen within the target halo can reach the density necessary for self-shielding.
The ``lateUVB'' and ``standard'' simulations show a similar behaviour before $z=5.5$.
In both cases, the filament remains neutral until the end of reionization with a cool core with temperature below $1000$\,K.
The ``standard'' simulation shows a stronger ionization in the upper left corner at $z=7$ and $z=9$, which is a hint for the effect of the additional external radiation. At $z=5.5$, the filament becomes instantaneously ionized and heated in the ``lateUVB'' case, while in the ``standard'' case, the ionization occurs gradually.
By $z=3$, the target halo still hosts significantly more neutral and cold gas in the ``standard'' simulation compared to the other two simulations.
The gas density, which we do not show in this figure, exhibits a similar pattern as the temperature, with a significantly smoother structure in the ``UVB'' simulation at $z \geq 6$ than in the other simulations.

\subsection{Baryon mass fraction}
The ionization and subsequent heating of the halo's neighbourhood, especially of the filament, both reduce the gas inflow and can unbind already accreted gas.
To analyze this loss of baryons due to external radiation, we show in \cref{fig:baryonFraction} the baryon mass to halo mass ratio $M_b / M_{\rm halo} = \left(M_\star + M_{\rm gas} \right) / M_{\rm halo}$ as a function of halo mass at different redshifts.
We note that this ratio is typically dominated by the gas mass at high redshift, as it is in our case.
To increase our sample size in the low halo mass regime, we take into account all haloes that only contain high-resolution particles.
To increase the sample on the higher mass end, we consider the five snapshots closest to the target redshift, resulting in a time range of around $40$\,Myr.
For each redshift, we additionally calculate the corresponding distributions of the L4N512 box from the \thesanHR simulation suite \citep{borrow2023thesan}, which uses the original \thesan model with an effective equation of state model for the ISM. 
The box was simulated with two different treatments of the radiation field: In the standard run, self-consistent RT was used, while in the UVB run, the same uniform UVB as in our ``UVB'' simulation set was used. We note that our ``UVB'' run still uses RT for local sources, so the radiation field, in general, should be stronger in our simulations. 
\thesanHR used a 10 times worse mass resolution per particle but covered a full cosmological box, which hosts more massive haloes than the zoom-in regions of our study.
Due to the small box with side length $5.9\,\mathrm{cMpc}$, reionization did not finish at $z=5$ in the simulation with radiative transfer.

Above $z=11$, the baryon fraction for most haloes is very close to the cosmic baryon fraction of $\approx 0.157$, with a slight suppression above $10^7 \msun$ due to previous stellar feedback and longer cooling times, which leads to a stronger gas pressure. 
After the onset of the UVB in the ``UVB'' set, haloes with $M_{\rm halo} < 10^8\msun$ lose a significant amount of their baryon content (see e.g. $z=9$).
In contrast, haloes above $10^9 \msun$ can retain most of their baryons even down to $z=3$, consistent with the picture that only haloes with a virial temperature (Eq.~\ref{eq:virialTemperature}) above $\sim10^4$\,K are able to keep ionized gas.
The ``standard'' and ``lateUVB'' simulation sets generally show similar behaviour. 
Before reionization, most haloes are able to retain the majority of their gas component.
However, after being reionized, haloes below $10^8\msun$ quickly lose a significant amount of their baryonic component until $z=5$. 
In the ``standard'' case, this transition is smoother due to the inhomogeneous reionization time; the first differences between the two sets are noticeable at $z=6.5$. In the ``lateUVB'' simulation set, the transition occurs for all haloes at $z=5.5$.
Only haloes above $10^9\msun$ are able to retain a significant amount of their baryons until $z=3$.
At this later time, all simulation sets show a similar distribution of baryons.

The suppression of the baryon fraction for haloes above $10^7 \msun$ already at $z=11$ is consistent with the  \thesanHR suite, though in their case, it is for all redshifts stronger.
This can be caused by our different galaxy formation models, especially the complete lack of local radiation feedback in their simulation with UV background.
After the onset of the uniform UVB, the baryon loss in smaller haloes is consistent with our ``UVB'' set.
In their simulation with radiative transfer, the baryon loss is weaker at $z=5$, which is a direct consequence of the incomplete reionization due to their small box size.
\cite{Wu2019a}, which uses the same RT solver as \thesan but the Illustris galaxy formation model, found at $z=7$ only a small baryon loss for haloes with masses below $10^9\msun$, which is consistent with our results. After reionization at $z=5$, their results are consistent with our ``lateUVB'' and ``standard'' sets, although their resolution did not allow them to resolve the full transition to haloes with a baryon fraction below 3\%.
We also compare our results at $z=11$ with those from \cite{Chen2014} (Rarepeak simulation, we use the values from $z=15$), which shows a suppression of the baryon fraction in haloes of the total mass of around $10^8\msun$.

We note that especially haloes with total mass below $10^7\msun$ can be affected at high redshift by processes we do not model, such as the cosmological streaming velocity between baryons and dark matter \citep{Long2022} or an Xray background \citep[e.g.][]{Jeon2014}.

\begin{table}
	\centering
	\begin{tabular}{c|cc|cc|cc} 
  \hline

  z & $a_{\rm std}$  & $b_{\rm std}$ &
  $a_{\rm UVB}$  & $b_{\rm UVB}$  &
  $a_{\rm lateUVB}$  & $b_{\rm lateUVB}$\\
  \hline
 11 & 0.29 & 7.75 & 0.25 & 7.78 & 0.26 & 7.76\\
 9 & 0.30 & 7.76 & 0.36 & 7.25 & 0.30 & 7.79\\
 7 & 0.47 & 7.66 & 0.43 & 7.59 & 0.42 & 7.74\\
 5 & 0.39 & 7.14 & 0.44 & 7.94 & 0.49 & 7.50\\
 3 & 0.43 & 7.32 & 0.58 & 8.33& 0.53 & 7.77\\
\hline
 \end{tabular}%
    \caption{The fitting parameters $a$ and $b$ as a function of redshift for the stellar occupation fraction as a function of halo mass (Eq.~\ref{eq:OF}) for the three different sets of simulations (std = ``standard'').
    We employ a least squares fitting method and observe that the fitting functions visually match the simulation results shown in \cref{fig:stellarFraction} almost perfectly.}
    \label{tab:fits} 
\end{table}
\begin{figure*}[
    \centering
    \includegraphics[width=1\linewidth]{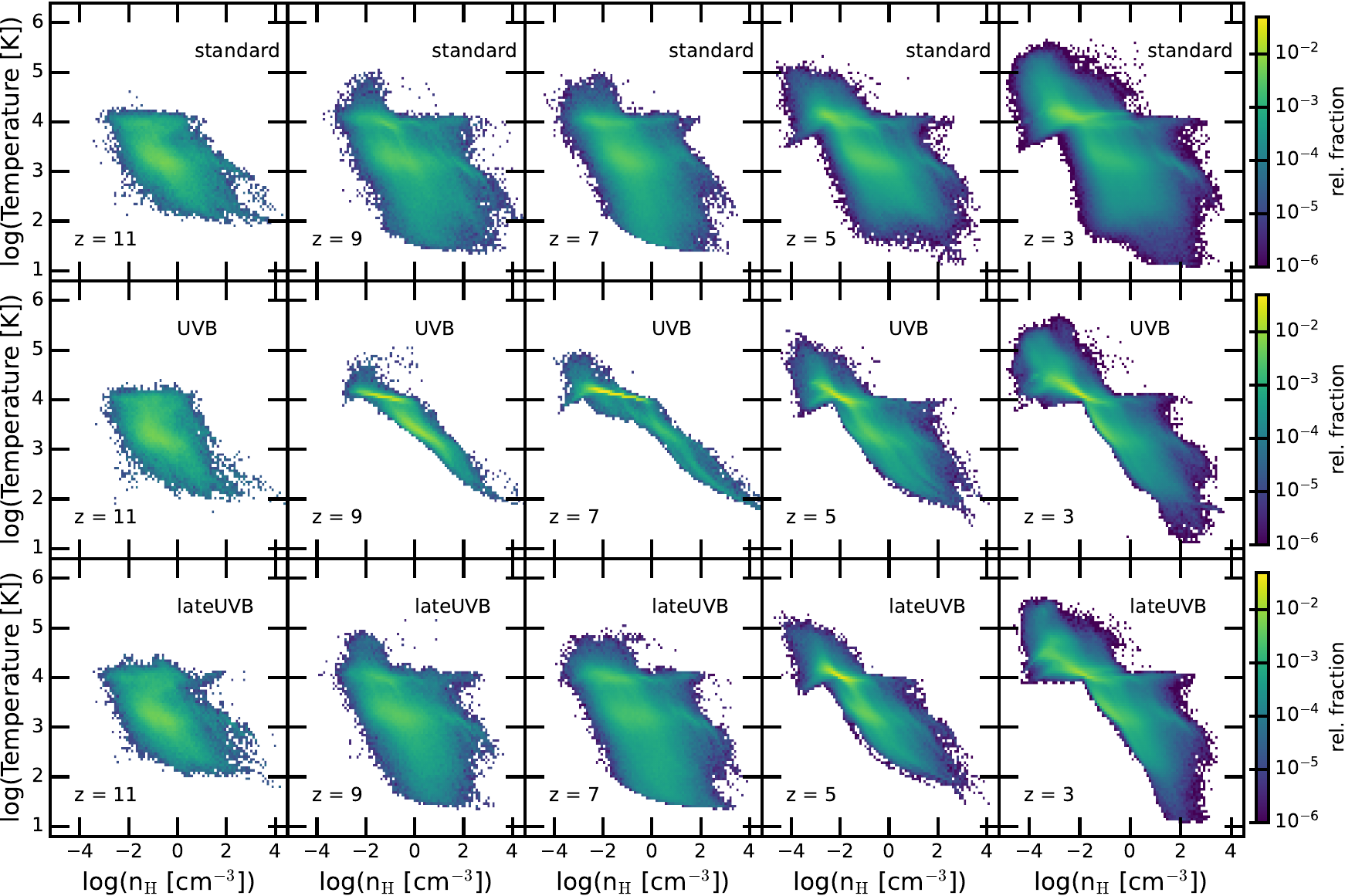}
    \caption{Phase space diagram at five different redshifts for gas cells bound to central subhaloes. For each physics modification and redshift, we use all gas cells associated with any of the seven central subhaloes of the target haloes in our sample and calculate the relative fraction of each bin.
    The first row shows the results with the radiation background from the \thesanone simulation, the second one with the uniform UVB starting at $z=10.6$, and the last one with the uniform UVB starting at $z=5.5$. 
    Before the onset of the external radiation field at $z=11$, all simulations show very similar behaviour. The haloes are still quite small, and no gas significantly hotter than $10^4$\,K is bound to the central subhalo.  
    At $z=9$, in the ``UVB'' simulation set, low-density and cool gas is not self-shielded, resulting in ionization and heating. Consequently, the phase space distribution becomes very narrow. 
    A similar pattern is observed at $z=5$ in the ``lateUVB'' simulation set, which also looks very similar at $z=3$ to the ``UVB'' set. In contrast, the ``standard'' simulations successfully retain some low-density and cold gas that remains shielded from the background radiation. 
    Further discussion can be found in \cref{sec:UVBGasPhase}.
    }
    \label{fig:phase_diagram_summary}
\end{figure*}

\subsection{Stellar-mass-to-halo-mass ratio}
Removing baryons through photoionization also suppresses star formation in smaller haloes. 
This becomes obvious in the evolution of the stellar mass to halo mass ratio and the occupation fraction, which is defined as the fraction of haloes containing at least one star particle, as shown in \cref{fig:stellarFraction} for our simulations.
We note that this definition of the occupation fraction depends on the mass resolution of stellar particles as long as not individual stars are resolved. 
We discuss in \cref{app:resolution} the results for alternative definitions of the occupation fraction using a fixed stellar mass threshold.
Before the onset of a significant external ionizing field ($z=11$), all simulation sets show a similar star formation with a threshold halo mass of around $10^8\msun$, above which most haloes contain at least one single star particle. 
This halo mass is approximately equivalent to the atomic cooling limit \citep{sutherland1993cooling}, which corresponds to the halo mass at which cooling from atomic hydrogen becomes important, leading to efficient star formation.
From $z=11$ to $z=9$, the occupation fraction for the ``UVB'' simulation set shifts to lower halo masses.
This is caused by increased star formation in haloes with mass $\approx 10^7\msun$ directly after the onset of the UVB.
These haloes typically only form a single star particle, which could be caused by more efficient cooling through a higher free electron density by the UVB.
When the UVB becomes stronger, haloes with a total mass below $10^9\msun$ become quenched.
They continue to accrete dark matter, which shifts the transition in the occupation fraction to larger halo masses already at $z=7$.
The  ``standard'' and  ``lateUVB'' sets evolve similarly until $z=7$.
After the onset of the radiation background in all simulations ($z=5$), the ``standard'' simulation set continues to form more stars in haloes below $10^8 \msun$ in comparison to the ``lateUVB'' simulation.
This becomes even more obvious at $z=3$ and could be explained by the more accurate self-shielding in the ``standard'' case, which not only considers the gas density but also performs radiative transfer for the external radiation. 
We compare for $z\geq 5$ also our results with those from the \thesanHR suite, which shows a similar suppression of star formation and occupation fraction below $10^9 \msun$. However, their worse mass resolution does not allow for a quantitative comparison at halo masses below $10^8\msun$, at which point their haloes contain at most one single star particle.
For $z=11$ we also compare with \cite{Chen2014} and \cite{shea2015}, which show a similar suppression of star formation in haloes with masses below $10^8\msun$.
We did not compare our results with the UNIVERSEMACHINE \citep{Behroozi2019}, since all of our haloes are deep in the regime in which the UNIVERSEMACHINE fits have to be extrapolated, as their fits are only valid for halo masses above $10^{10.5}\msun$.

The stellar occupation fraction provides a redshift-dependent cut-off for the halo mass, below which haloes are not able to host galaxies. 
This is an important ingredient for semi-analytical models \citep[e.g.][]{Bose2018}.
We fit our simulation results for the occupation fraction using the following equation:
\begin{equation}
\mathrm{OF} = \frac{1}{2} \left[\tanh\left(\frac{\log_{10} (M_{\rm halo}/\msun)-b}{a}\right) + 1\right] .
\label{eq:OF}
\end{equation}
We introduced the redshift dependent constants $a$ and $b$ given in \cref{tab:fits}.
$a$ corresponds to the width of the transition while $b$ gives the halo mass with $\mathrm{OF} = 0.5$.
For the ``UVB'' set, we observe a drop in the cut-off mass after the onset of the radiation background, which, as we discussed above, we associate with a short increased star formation in smaller haloes due to the increased free electron density.
After reionization, these haloes continue to accrete mass, especially dark matter, which increases $b$.
$b$ additionally increases due to newly formed haloes which are not able to form stars.
We observe a similar trend in the other two simulation sets:
Before the onset of a strong radiation background, $b$ remains relatively constant, though it slightly drops at $z=7$ in the ``standard'' set, a hint to a slowly rising radiation field.
At $z=5$, the cut-off reaches its minimum value before beginning to increase again.

\section{The impact of external radiation on the gas}
\label{sec:UVBGasPhase}

\begin{figure*}
    \centering
    \includegraphics[width=1\linewidth]{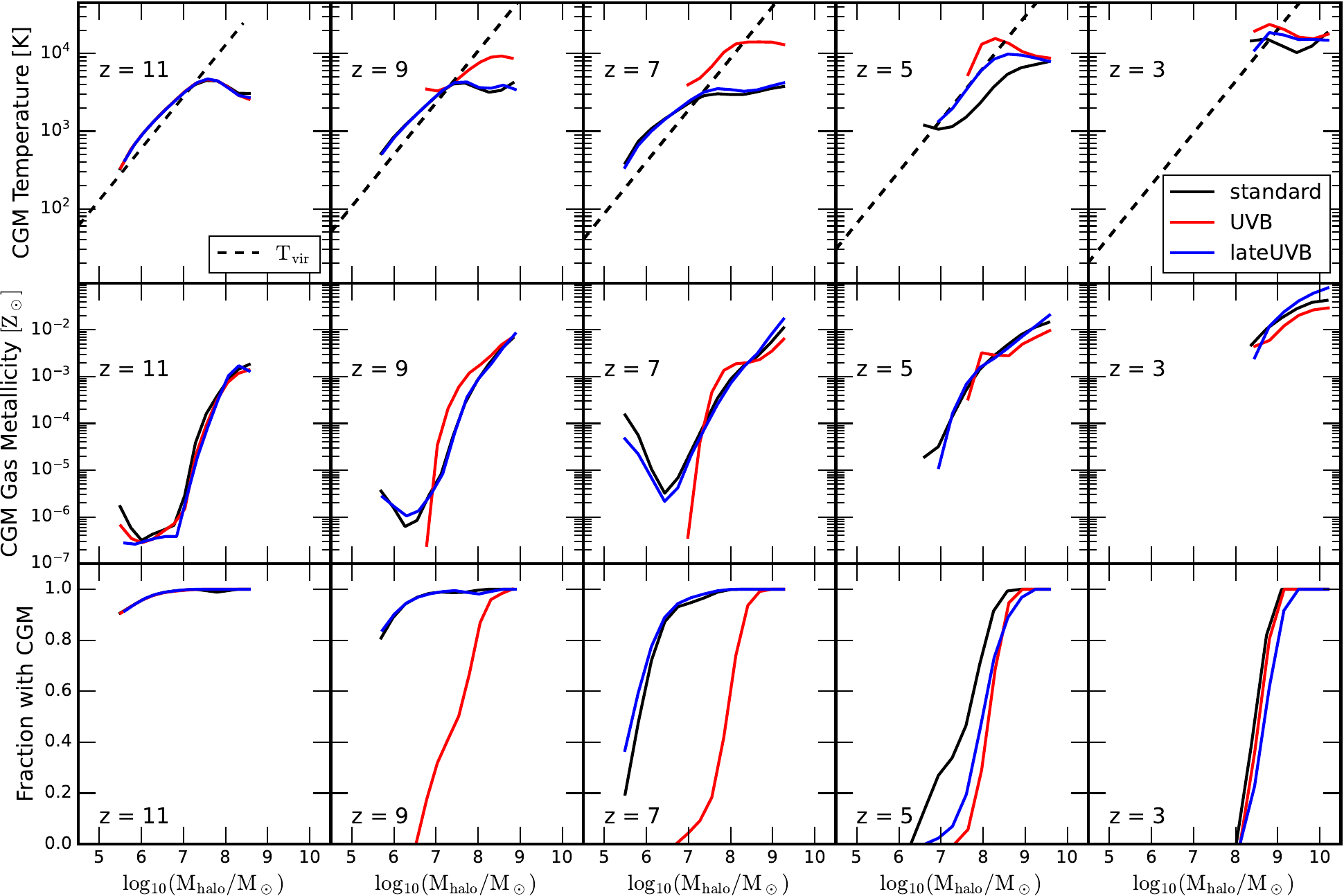}
    \caption{The temporal evolution of averaged properties of the CGM for all high-resolution, central galaxies is compared between three sets of simulations, each differing in their model of the large-scale external radiation field. 
    Top panels: The mass-weighted mean CGM temperature as a function of halo mass. The dashed line indicates the virial temperature, as defined in Eq.~(\ref{eq:virialTemperature}) \ assuming fully ionized gas.
    Before the activation of the external radiation field, the virial temperature accurately describes the results of the simulations for smaller haloes. Above $10^7 \msun$, cooling occurs, leading to star formation.
    The uniform UVB quickly heats the gas to around $10^4$\,K.
    Middle panels: The mass-weighted mean metallicity normalized by solar metallicity as a function of halo mass. 
    Star formation in haloes above $10^7\msun$ results in chemical enrichment and more effective cooling. 
    The reduced star formation in the ``UVB'' simulation set results in slightly lower metallicities at $z=3$.
    Lower panels: Fraction of haloes that have at least one gas cell bound to their central subhalo. Before reionization, most haloes are able to maintain a CGM.
    However, the radiation background fully ionizes most haloes below $10^8\msun$. This ionization process occurs more slowly in the ``standard'' simulation set due to the finite time it takes for radiation to travel through the zoom-in region, as well as the more effective self-shielding.
     For better statistics we use the five snapshots closest to the target redshift, resulting in a time range of around $40$\,Myr.
   }
    \label{fig:cgm_properties}
\end{figure*}
\subsection{Phase space distribution}
The multiphase gas model of the \thzoom project allows for a detailed study of the gas structure within haloes, in contrast to the \thesan project, which used the equation of state model from \cite{springel2003cosmological}.
The latter one leads to an artificial pressure boost for densities exceeding $n_\mathrm{H} = 0.13~\mathrm{cm}^{-3}$. In contrast, the phase space diagram in \cref{fig:phase_diagram_summary} shows that the gas in the \thzoom galaxies possesses a significantly more complex structure with a multiphase distribution.
We binned all gas cells bound to the central subhaloes of the seven target haloes at different redshifts and subsequently normalized the histogram.
Before external reionization, we find at $z=11$ a similar phase structure for all models, with a minimum temperature of approximately $100$\,K. 
This is slightly lower than the minimum temperature reached by primordial gas through cooling by molecular hydrogen \citep[$\approx 200$\,K,][]{greif2015numerical}, indicating some presence of metal line cooling. There is also an upper-temperature limit of around $2\times 10^4$\,K, corresponding to the maximum virial temperature of our haloes at this high redshift. Even warmer gas would be unbound from the subhalo.
There is an abundance of cold and low-density gas as well as star forming gas with densities reaching $n_\mathrm{H} = 10^4~\mathrm{cm}^{-3}$.
At $z=9$, significantly lower temperatures are reached in the ``standard'' and ``lateUVB'' simulation sets due to previous metal enrichment, which allows efficient low-temperature metal line cooling. The haloes have also grown and are able to keep a bound low-density ($n_\mathrm{H} < 10^{-1}~\mathrm{cm}^{-3}$), hot component ($T > 10^{4.5}~\mathrm{K}$) as well as a dense and warm component in \hii regions ($n_\mathrm{H} > 10~\mathrm{cm}^{-3}$, $T \approx 10^{4}~\mathrm{K}$). 
The uniform radiation background in the ``UVB'' simulation set rapidly changes the phase-space structure of the gas that remains bound to the halo. Low-density gas is heated, forming a homogenous warm hot medium with $T\approx 10^4\,K$ and $-3 < \log\left(\rm n_H / cm^{-3}\right) < 0$.
The radiation feedback from the smaller stellar component (see \cref{fig:stellarFraction}) is not strong enough to
create a warm and dense phase at $z=9$. The lowest temperature reached in this simulation set is higher than for the other two, which can be explained by the absence of self-shielding for $z > 6$.
After the onset of self-shielding following the simplified description from \cite{Rahmati2013}  at $z=6$, the phase space diagram becomes less narrow.
The ``standard'' and ``lateUVB'' sets continue to evolve similarly until the onset of the uniform UVB radiation in the later set at redshift $z=5.5$. 
By $z=5$. the ``lateUVB'' simulation set has lost its low-density, cold gas phase and its phase space structure becomes more similar to the ``UVB'' set. In contrast, the more self-consistent self-shielding based on radiative transfer in the ``standard'' set allows this phase to exist until $z=3$.
At $z=3$ metal line cooling allows in all simulation sets gas to cool almost down to the redshift-dependent temperature floor $\rm T_{\rm floor}(z) = \left(1+z\right) T_{\rm floor}(0)$, set by the cosmic microwave background radiation (CMB)  with the present temperature $\rm T_{\rm floor}(0) = 2.726~\mathrm{K}$ \citep{fixsen2009temperature}.
We note that in our simulations, the actual temperature floor is set to $12~\mathrm{K}$, slightly higher than the one set by the CMB at $z=3$. 

\subsection{Impact on the CGM}

External radiation must cross the circumgalactic medium (CGM) before interacting with the star-forming interstellar medium.
Any gas that is being accreted into the centre of the galaxy must also first cross the CGM, which serves as a gas reservoir. 
To define the CGM in our simulations, we use the definition from \cite{borrow2023thesan}: We select all central subhaloes and consider only the gas that is bound to them.
Additionally, we calculate the stellar half-mass radius $R_{1/2, *}$ of the subhalo and exclude all gas cells within a sphere of radius $2R_{1/2, *} $. This condition removes the ISM; however, since the criterion is purely spatial, it still includes recently accreted high-density gas.
Gas significantly above the virial temperature is not bound to the central subhalo and is therefore ignored in our analysis.
According to the stellar occupation fraction shown in \cref{fig:stellarFraction}, the smallest resolved haloes in our simulations do not contain any stellar particles, which does not allow us to define $2R_{1/2, *} $. In this case, we set $R_{1/2, *} =0$, which means we consider all gas bound to the central subhalo.

We show in \cref{fig:cgm_properties} the mass-weighted mean temperature and gas metallicity of the CGM for all central subhaloes of haloes that contain only high-resolution particles.
We normalize the metallicity by the solar metallicity $Z_\odot = 0.0127$ \citep{wiersma2009chemical} and calculate the fraction of haloes that contain at least one gas cell within their CGM. 
At high redshift ($z=11$), all simulation sets show a similar temperature distribution, which is described by the virial temperature defined in Eq.~(\ref{eq:virialTemperature}) for halo masses below $2\times 10^7 \msun$.
In this mass range, haloes do not form any stars, and the gas remains nearly pristine. Primordial cooling is inefficient in removing the heat produced by accretion.
In contrast, larger haloes can form stars, which enrich the halo with metals. The more efficient metal line cooling reduces the temperature of the CGM.
At high redshift almost all haloes are able to retain a significant gas reservoir in their CGM due to the absence of external radiation sources.

The uniform radiation background in the ``UVB'' simulation set is capable of fully ionizing most of the CGM, resulting in heating of the CGM to around $10^4$\,K after $z=10.6$. 
Haloes with virial temperature below this threshold are not able to keep their CGM gravitationally bound, which results in haloes below $10^8\msun$ losing their CGM by $z=9$.
For the other two simulation sets, the external radiation field has a minor impact at $z=9$.
At $z=7$, some haloes below $10^7 \msun$ start to lose their CGM due to radiation feedback from nearby high-resolution haloes in the zoom-in region in both the ``standard'' and ``lateUVB'' simulation sets. 
At $z=5$, all simulation sets are fully ionized, and haloes below $10^9\msun$ start to lose their CGM. This process is faster in the ``lateUVB'' set compared to the ``standard'' set.
In the former one, the radiation background is directly applied to all gas cells below the density threshold for self-shielding.
Conversely, in the ``standard'' set, the radiation must first travel through the high-resolution region, and low-density areas can be self-consistently shielded, as seen in  \cref{fig:phase_diagram_summary}.
By $z=3$, the CGM fraction is similar across all simulation sets. However, the ``UVB'' set has a slightly lower metallicity in the CGM due to the reduced stellar feedback in the past as can be seen in the stellar mass-halo mass relation in \cref{fig:stellarFraction}.

The loss of the CGM due to the external radiation field is also consistent with the baryon fraction shown in \cref{fig:baryonFraction} and the suppressed star formation after reionization in haloes with total masses below $10^9\msun$ as discussed for the stellar occupation fraction in \cref{fig:stellarFraction}.
This loss could be reduced by more efficient cooling through early metal enrichment, which could be caused by Population III stars.
As discussed in a companion paper \citep{ZierPopIII}, we do not include the higher metal yields and energy production from these stars \citep{takahashi2018stellar}.
However, this enrichment would have to be significant since for metallicities $Z \ll Z_\odot$, efficient primordial cooling dominates at $T\approx 10^4$\,K \citep{Ploeckinger2020}.

\begin{figure}
    \centering
    \includegraphics[width=1\linewidth]{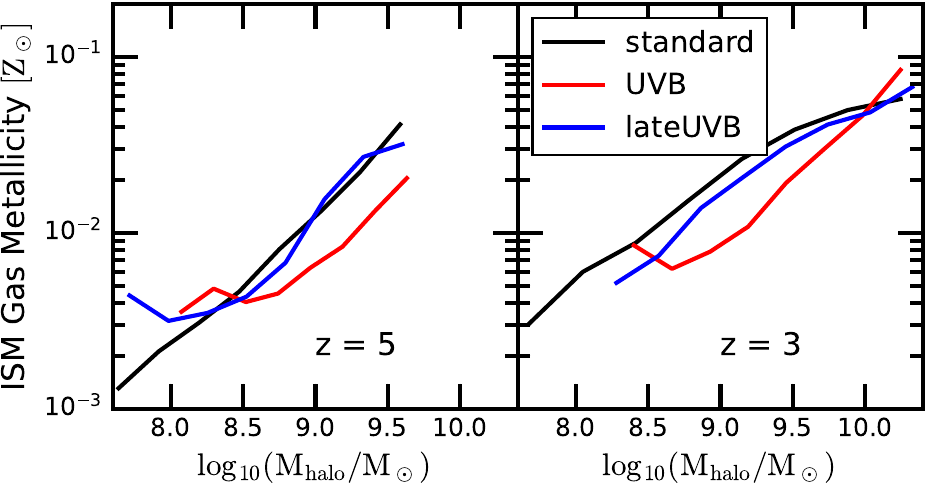}
    \caption{The mass-weighted mean gas metallicity within twice the stellar half-mass radius $2 R_{1/2, *} $ around the centre of the central subhalo (defined as the ISM) at $z=5$ and $z=3$ as a function of the halo mass. Our analysis only includes central subhaloes within haloes composed entirely of high-resolution particles.
    The ``UVB'' set shows a lower metallicity in subhaloes with masses below $10^{10}\msun$. This decrease in metallicity is attributed to suppressed star formation in the progenitors of these haloes during the reionization era, which also leads to the reduced stellar-to-halo mass ratio seen in \cref{fig:stellarFraction}.
    We use the five snapshots closest to the target redshift for better statistics, resulting in a time range of around $40$\,Myr.}
    \label{fig:ism_metallicity}
\end{figure}

\begin{figure*}
    \centering
    \includegraphics[width=1\linewidth]{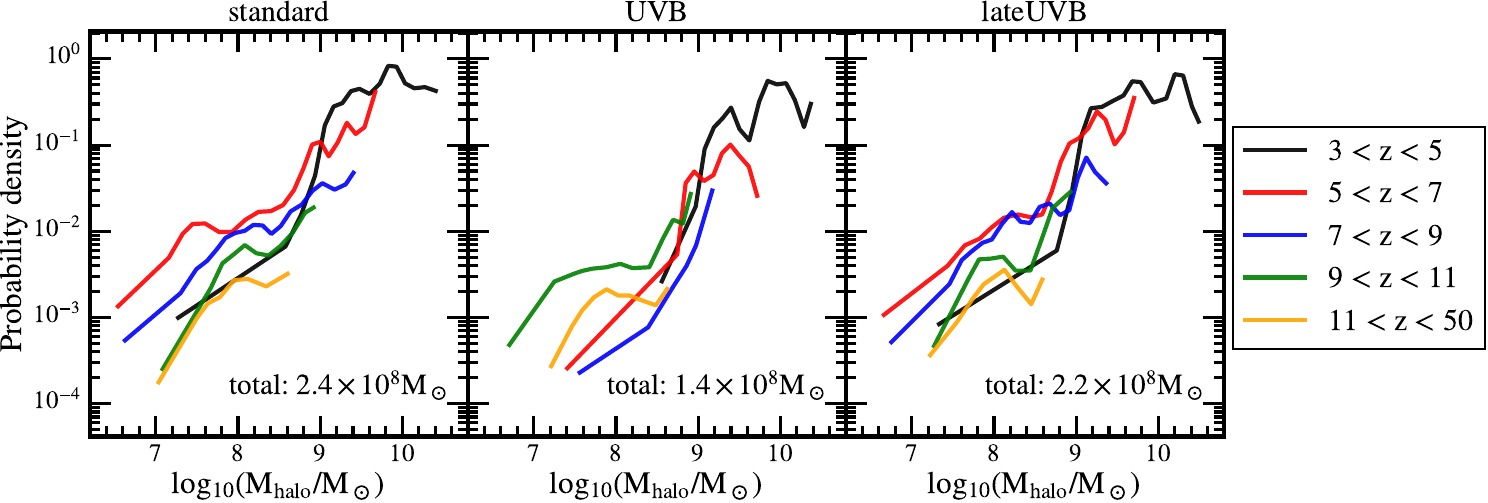}
    \caption{The distribution of birth halo mass for all star particles that can be found in a high-resolution halo at redshift $z=3$ in our simulation sets. 
    All distributions are normalized by the total stellar mass formed in the ``standard'' set at $z=3$ ($2.4\times 10^8 \, \msun$). By integrating this function over the entire halo mass range, we obtain values of 1, 0.58, or 0.92, respectively.
    We show the distribution for stars born in five different redshift ranges and also provide the total stellar mass formed across all redshifts. 
    The radiation background suppresses star formation in haloes with masses below $10^9\msun$, with an earlier suppression observed in the ``UVB'' set due to the earlier initialization of the UVB.
    This earlier suppression results in approximately 50\% fewer stars at $z=3$ compared to the ``standard'' set.
    }
    \label{fig:birth_halo_distribution}
\end{figure*}

\begin{figure*}
    \centering
    \includegraphics[width=1\linewidth]{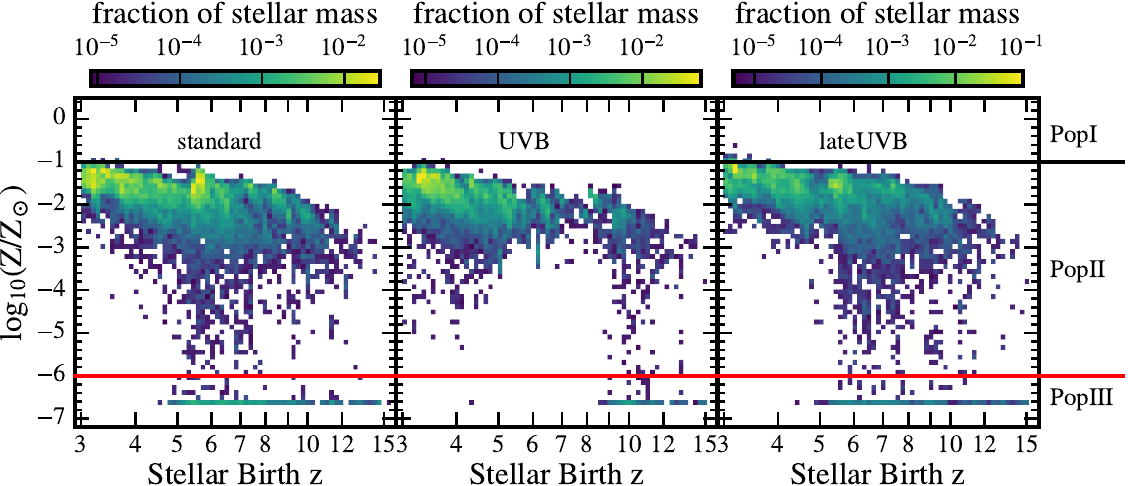}
    \caption{We show the distribution of stellar particles weighted by their birth mass from the seven target haloes within the zoom-in regions at redshift $z=3$ as a function of the stellar metallicity and their birth redshift. 
Each histogram corresponds to different simulation sets, which vary in their implementation of the external radiation field.
    The histograms are normalized by the total stellar mass, and we also indicate metallicity thresholds for various stellar populations. Both the ``standard'' and ``lateUVB'' simulation sets show a continuous formation of Pop III stars down to $z=5$. In the ``lateUVB'' scenario, the minimum metallicity required for star formation increases more rapidly after the end of reionization. 
    The ``UVB'' set shows a markedly different distribution: Pop III  star formation ends at $z=9$, and during the EoR, there exists a lower metallicity threshold of approximately $10^{-4}\,Z_\odot$. Below $z=5$, this metallicity threshold gradually decreases. 
    We refer to \cref{sec:UVBStellarEvolution} for a more detailed discussion.}
    \label{fig:age_metallicity_stars}
\end{figure*}

\subsection{Impact on the ISM}
The ISM serves as a birthplace for stars, establishing a direct connection between the properties of gas and those of stars.
To avoid contributions from the halo, we define the ISM as all bound gas contained within a sphere of radius $2 R_{1/2, *} $ around the centre of the central subhalo.

As discussed for the CGM before, this definition becomes problematic in very small haloes that do not have a well-resolved stellar population.
We will concentrate in the following on galaxies that host at least one stellar particle as well as one bound gas cell, which results in a lower halo mass limit of $\approx 10^8 \msun$ after reionization.
In \cref{fig:ism_metallicity}, we show the mass-weighted mean metallicity in the ISM at redshifts $z=5$ and $z=3$ as a function of the halo mass.
Haloes with masses below $10^{10}\msun$ show a significantly lower metallicity in their ISM in the ``UVB'' simulation, which can be explained by the suppressed star formation at higher redshift in small haloes. 
More massive haloes are less affected by external radiation since their progenitors were massive enough to form stars during reionization.
The ``lateUVB'' set shows at $z=3$ also slightly lower metallicity than the ``standard'' set, which could be explained by the slightly more efficient suppression of star formation in that scenario.

\section{Influence of external ionization on stellar properties}
\label{sec:UVBStellarEvolution}

As discussed in the previous sections, gas properties are sensitive to the implementation of the radiation background during reionization.
These differences become smaller over time when all haloes are impacted by the external UV radiation, as e.g. seen at $z=3$ in \cref{fig:cgm_properties}.
Stars that formed during reionization retain their initial metallicity and age and, therefore, can act as a memory of the past state of the gas.

\subsection{Surpression of star formation by external radiation}
By tracing back all stars that are within high-resolution haloes at $z=3$ to their birth haloes in the first snapshot after their formation, we are able to analyze the importance of different halo mass regimes on the total star formation rate.
We show in \cref{fig:birth_halo_distribution} the distribution of the birth halo mass for stars born during five different redshift ranges. 
Star formation in haloes with masses below $10^9\msun$ is almost fully shut down after the onset of the UVB in the ``UVB'' set, consistent with the loss of the CGM in these haloes.
Larger haloes are able to continue forming stars until the end of the simulation at $z=3$, albeit at a lower rate than in the other two simulations before the end of the EoR.
Both the ``standard'' and ``lateUVB'' simulation sets show continuous star formation until $z=5.5$ in haloes with masses even below $10^7\msun$.
However, the completion of reionization suppresses star formation in haloes below $10^9\msun$, while larger haloes are able to continue forming stars.
The decreased star formation in the ``UVB'' case results in nearly 50\% less total stellar mass at $z=3$, though we note that our sample only includes one halo above $10^{10}\msun$.
Haloes above this threshold dominate the star formation rate density below $z=6.5$ in the \thesanHR box, with more massive haloes dominating in the \thesanone box \citep[e.g.][]{Yeh2023}.
Therefore, the overall suppression of star formation might be less significant when considering a full cosmological box compared to our sample.

\subsection{Impact of external radiation on stellar metallicities}
Suppressing star formation in low-mass haloes also affects the metallicity of newborn stars, as we can see in \cref{fig:age_metallicity_stars}. 
We considered all stars belonging to one of the seven target haloes of our zoom-in regions at $z=3$ and created a two-dimensional histogram that shows the relationship between stellar metallicity and birth redshift.
In a companion paper \citep{ZierPopIII}, we analyze this relation in greater detail, specifically focusing on low metallicity stars for the ``standard'' simulation set.
As discussed in the companion paper, we define three distinct stellar populations based on metallicity: Pop I stars are defined as having $Z > 0.1\,Z_\odot$, Pop III stars have $Z < 10^{-6}\,Z_\odot$ and Pop II stars fall in between. 
Our upper limit for Pop III stars is rather conservative, as other studies have used limits of $10^{-4}\,Z_\odot$ \citep{jaacks2019legacy, liu2020gravitational, venditti2023needle} or $10^{-5}\,Z_\odot$ \citep{Ricotti2016, sarmento2018following,sarmento2022effects, Brauer2024}. 
We selected our conservative threshold because we do not include an explicit model for Pop III stars, particularly regarding their higher metal yields and energy injection. As a result, our criterion primarily identifies stars that formed from pristine gas as Pop III stars.
The transition criteria from a top-heavy Pop III IMF to a standard Pop II IMF is still debated \citep[e.g.][]{maio2011interplay, chiaki2014dust, chon2021transition, chon2022impact,chon2024impact}.

\begin{figure}
    \centering
    \includegraphics[width=0.95\linewidth]{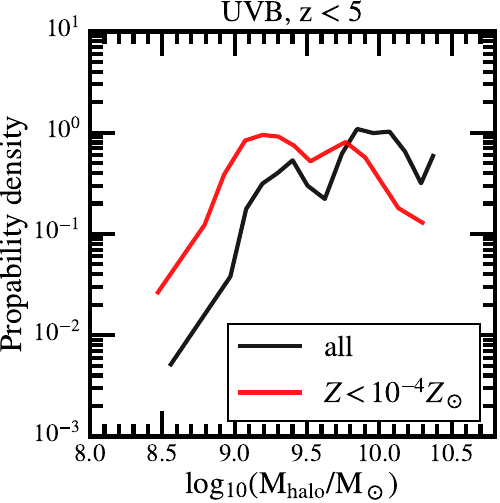}
    \caption{The probability density of the birth halo mass distribution for stellar particles formed after redshift $z=5$ in the ``UVB'' set. 
    The results are shown for all stars (black line) and for stars with metallicities below $Z= 10^{-4}\,Z_\odot$ (red line).
    Stars with low metallicities are more likely to be born in smaller haloes with masses below $10^{9.5} \msun$.
    This is only slightly above the halo mass threshold below which no stars can be formed after reionization.
    The progenitors of these haloes showed, on average, reduced star formation and metal pollution.
    This can explain why these metal-poor stars cannot form after $z=5$ in the other simulation sets with later reionization.
}
    \label{fig:birth_halo_distribution_low_red}
\end{figure}

The ``standard'' and ``lateUVB'' show a similar distribution, with Pop III star formation continuing until $z=5$. As we discuss in the companion paper, this happens mainly in so-called minihaloes with masses below $10^8\msun$ and is also compatible with other simulations \citep[e.g.][]{jaacks2018baseline,jaacks2019legacy, liu2020gravitational,liu2020did, venditti2023needle, borrow2023thesan}. 
Reionization suppresses star formation in these small haloes, and later on, only higher metallicity stars can form. 
This suppression is stronger in the ``lateUVB'' simulation set.
The ``UVB'' set shows at around $z=8.5$ a jump in the minimum metallicity associated with the growing UVB.
However, the minimum metallicity starts to decline again below $z=5$, allowing for the formation of stars with metallicities below $10^{-4}Z_\odot$ down to $z=3$.
To better understand the different behaviour at lower redshift between the different simulation sets, we show in \cref{fig:birth_halo_distribution_low_red} the birth halo mass distribution of stars born after $z=5$ in the ``UVB'' set. 
Low metallicity stars are born predominantly in haloes below $5\times 10^9 \msun$, which is close to the lower boundary of haloes being able to form stars after reionization.
In the ``UVB'' set, their main progenitor showed reduced star formation and, therefore, metal pollution.
Additionally, accreted minihaloes contain more pristine gas, as they did not form Pop III stars like those in the ``standard'' set.
The lower stellar metallicity in these small haloes is also compatible with the reduced ISM metallicity at $z=3$, as shown in \cref{fig:ism_metallicity}.

\section{Summary}
\label{sec:summary}
Regions can be ionized by both local and external sources.
Zoom-in simulations that do not explicitly model the external sources have to resort to simplified assumptions such as a redshift-dependent, uniform UV background.
In this paper, we compared three different treatments of the external UV radiation in zoom-in simulations,  utilizing seven regions corresponding to low-mass target haloes from the \thzoom suite. The models for external radiation are directly linked to different reionization histories:
The ``standard'' set employs the large-scale radiation field from the parent RHD \thesanone simulation as a boundary condition, which represents a patchy reionization process.
The ``UVB'' set utilizes a spatially uniform but time-varying UV background commonly used in cosmological simulations, activated at $z = 10.6$. This approach leads to flash reionization of the IGM, ending at $z = 10$.
The final ``lateUVB'' set also uses a uniform but time-varying UV background; however, it is activated at $z = 5.5$, which is approximately the end time of reionization in \thesanone.
The relatively high mass resolution of $6 \times 10^3 \msun$ per dark matter particle and $1.1\times 10^3 \msun$ for gas and stars allows us to resolve the smallest star-forming haloes with several hundred particles. This, combined with a self-consistent multi-phase ISM model and radiation transport, enables a detailed analysis of the effects of external radiation on galaxies and their stellar components.
We find that the timing of external reionization significantly impacts galaxies with halo masses below $10^9 \msun$. Due to the growth of these haloes over time and their merging with lower mass haloes, the timing of reionization can also affect the chemical distribution of newborn stars in larger haloes up to $10^{10} \msun$.
Specifically, we find:
\begin{enumerate}
    \item The uniform UV background results in nearly instantaneous reionization and heating of the IGM to $10^4$\,K. This process decreases the baryon fraction in haloes with masses below $10^9 \msun$ and removes all bound gas in most haloes below $10^8 \msun$.
    \item This loss of gas leads to a suppression of star formation in haloes below $10^9 \msun$ at the time of reionization of the specific region. Consequently, there is a reduction in stellar fractions in areas with earlier reionization, as well as a decrease in stellar occupation fractions. The suppressed star formation in progenitors results in lower metallicity of the interstellar medium (ISM) in galaxies with halo masses below $10^{10} \msun$, even at redshift $z=3$ for the ``UVB'' set.

        \item In the ``lateUVB'' and ``standard'' sets, Pop III stars continue to form in minihaloes until $z\approx 5$ (see \cite{ZierPopIII} for further discussion). 
    In contrast, these haloes are completely quenched in the ``UVB'' set, allowing star formation only from pre-enriched gas below $z=9$. The lack of enrichment in minihaloes leads to the formation of enriched but extremely metal-poor stars ($< 10^{-4}\,Z_\odot$) down to $z=3$, which are not observable in the other two sets.
    \item Galaxies with halo masses above $10^{10} \msun$ are less influenced by external reionization; they are capable of retaining a bound gas phase and maintaining continuous star formation. This is expected, as their virial temperature exceeds the temperature achieved through photonionization of hydrogen. However, due to the small size of the zoom-in regions, we are unable to extend our comparisons to larger halo masses.
    \item The self-consistent treatment of shielding in the ``standard'' set, by employing radiative hydrodynamics transfer for external radiation, enables the survival of a cold, low-density gas phase ($n_\mathrm{H} \approx 0.1~\mathrm{cm}^{-3}$, $T\approx 100$\,K). In contrast, using a purely density-based self-shielding criterion \citep[e.g.][]{Rahmati2013}, which is typically applied with a uniform UV background, results in the direct ionization of this phase.
\end{enumerate}
These results are qualitatively consistent with those from the \thesanHR project, which also found that galaxies with halo masses below $10^9 \msun$ experience quenching due to the UV background. 
\thesanHR finds a reduced baryon fraction for haloes at the high mass end already at $z=11$, which we attribute to differences in the galaxy formation model.
Similarly to our study, they observed continuous Population III star formation in minihaloes until their reionization and only a minor effect of the UV background on haloes exceeding $10^{11}\msun$. This consistency indicates that our results are robust against model variations, as the \thesanHR project employed a poorer mass resolution (8 to 64 times worse), used a simplified equation of state model for the interstellar medium, encompassed a full cosmological box rather than a zoom-in setup, and lacked a self-consistent external radiation field, relying instead on internal sources from a relatively small box. The highest resolution \thesanHR box probed a 40 larger volume at $z=8$ than all the high-resolution zoom-in regions analyzed in this paper combined, and its uniform volume provides a less biased perspective for the lowest-mass haloes. On the other hand, \thzoom probes the entire halo mass range from $10^6\msun$ to $10^{10} \msun$ more efficiently. In the future, we plan to run high-resolution small-volume simulations with both the \thesan and \thzoom galaxy formation models to statistically understand the systematic differences between models, overdensity, and reionization timing (\thesanVR).

Our results suggest that using a uniform UV background for high-resolution studies focused on early galaxy formation \citep[e.g. FireBOX][]{firebox} will affect galaxy properties, particularly their star formation history before $z=5.5$ , as well as the chemical distribution of newly formed stars down to at least $z=3$.
As demonstrated in \cite{ThesanEnrico}, for a baryon resolution of $\approx 5 \times 10^6 \msun$ per cell, the \thesan model yields results very similar to those obtained from the same simulation without radiative transfer but with a spatially constant UV background. With such low resolution, only massive galaxies above $10^{10}\msun$ can be adequately resolved. These haloes are less vulnerable to external radiation. 
However, these simulations will miss the pre-enrichment of gas within minihaloes at earlier times, potentially leading, as we illustrate in \cref{fig:age_metallicity_stars}, to the artificial formation of low metallicity stars down to $z = 3$.

Simulations designed to model galaxy formation during the epoch of reionization, or the formation of dwarf galaxies that produce most of their stars prior to reionization, require a realistic distribution of spatially non-uniform reionization redshifts for effective comparison with observations.
This can be achieved through full RHD simulation as \thesanone, although these require relatively large volumes to ensure a converged reionization history. 
Alternatively, one could use a spatially inhomogeneous, semi-numerical UVB \citep[e.g. in][]{bird2022astrid,ni2022astrid, Trac2022, Puchwein2023, Zhu2024}. This method is computationally less demanding than full radiative transfer simulations. While semi-numerical methods can be calibrated against full RHD simulations, they might not fully account for effects such as the shielding of low-density gas by higher-density regions. 

Simulations targeting larger galaxies that form the majority of their stars after reionization \citep[e.g. the EAGLE simulation forms more than 90\% of their stars between $z=2$ and $z=0$,][]{Crain2015}, typically do not have the resolution to accurately resolve minihaloes and their enrichment by Population III stars.
In such cases, a sufficiently high metallicity floor is necessary to prevent the formation of artificially low-metallicity stars at low redshift. Ideally, this metallicity floor should be spatially non-uniform and should take into account the local reionization redshift to effectively model the efficiency of metal enrichment within minihaloes \citep[see, e.g.][ for insights on the influence of different subgrid models for metal enrichment by Pop III stars on dwarf galaxies]{Gutcke2022PopIII}.
Moreover, the fraction of stars that are born within the main progenitor of a galaxy (in-situ fraction), particularly at high redshift, will be impacted by the absence of minihaloes or their artificial suppression in the presence of a uniform UV background.

\section*{Acknowledgements} 
The authors gratefully acknowledge the Gauss Centre for Supercomputing e.V. (\url{www.gauss-centre.eu}) for funding this project by providing computing time on the GCS Supercomputer SuperMUC-NG at Leibniz Supercomputing Centre (\url{www.lrz.de}), under project pn29we. 
Support for OZ was provided by Harvard University through
the Institute for Theory and Computation Fellowship.
RK acknowledges support of the Natural Sciences and Engineering Research Council of Canada (NSERC) through a Discovery Grant and a Discovery Launch Supplement (funding reference numbers RGPIN-2024-06222 and DGECR-2024-00144) and York University's Global Research Excellence Initiative. EG is grateful to the Canon Foundation Europe and the Osaka University for their support through the Canon Fellowship. WM thanks the Science and Technology Facilities Council (STFC) Center for Doctoral Training (CDT) in Data intensive Science at the University of Cambridge (STFC grant number 2742968) for a PhD studentship. XS acknowledges the support from the National Aeronautics and Space Administration (NASA) grant JWST-AR-04814. LH acknowledges support by the Simons Collaboration on ``Learning the Universe''.

\section*{Data Availability}
All simulation data, including snapshots, group, and subhalo catalogues and merger trees will be made publicly available in the near future. Data will be distributed via \url{www.thesan-project.com}. Before the public data release, data underlying this paper will be shared on reasonable request to the corresponding author.



\bibliographystyle{mnras}
\bibliography{main} 




\appendix
\section{Influence of resolution}
\label{app:resolution}
\begin{figure*}
    \centering
    \includegraphics[width=1\linewidth]{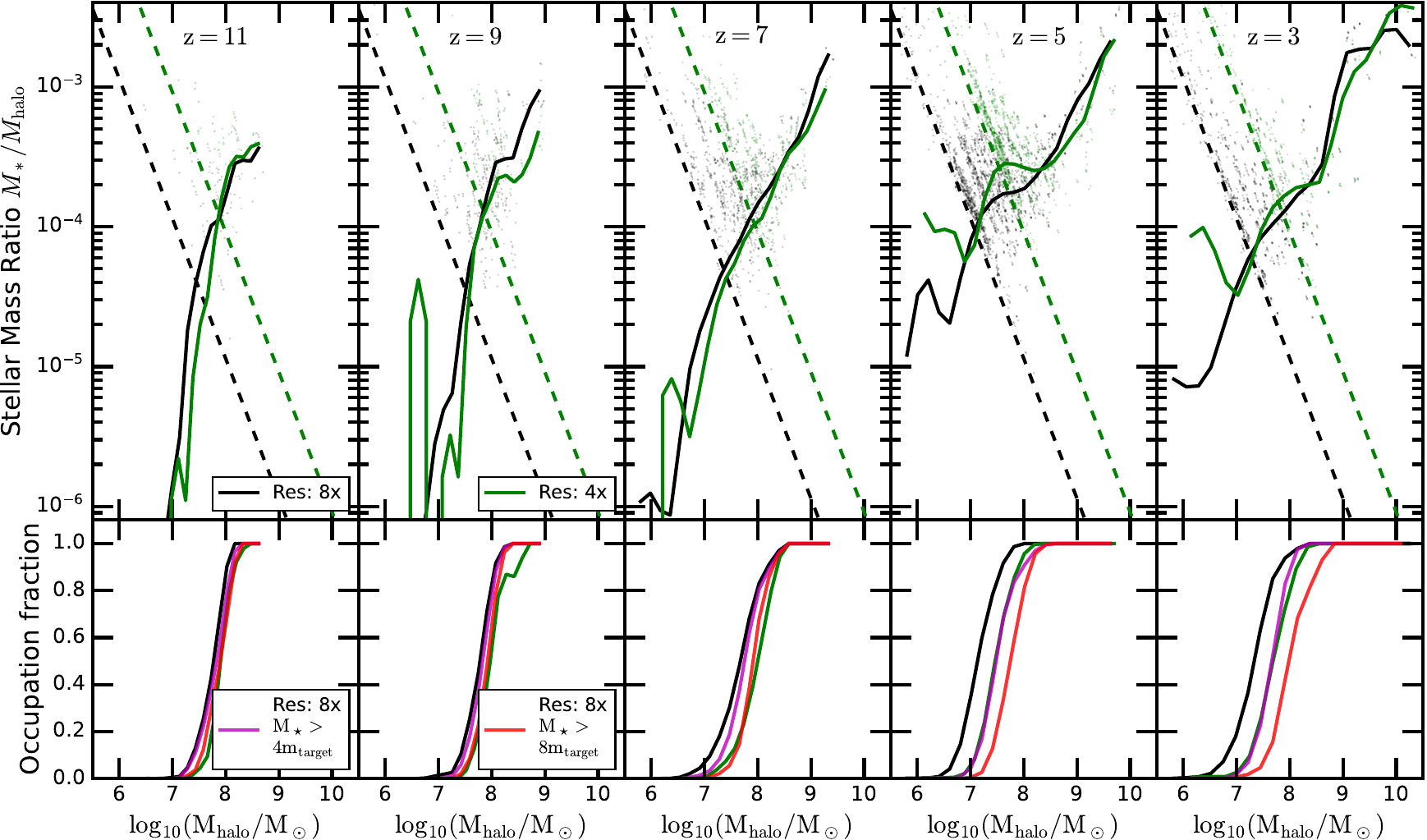}
    \caption{The evolution of the stellar mass to halo mass relation through cosmic time for the ``standard'' simulation set but two different resolutions. The zoom factor 8 (black line) was used in the main part of the paper, while the zoom factor 4 (green line) uses eight times worse mass resolution.
    The dashed lines represent the single-particle limit for both resolution levels. 
    Haloes above $10^8\msun$ are well resolved in both simulation sets and insensitive to changes in the resolution. 
    The occupation fraction, defined as the number of haloes containing at least one stellar particle, is shifted to larger masses for lower resolution.
    To compensate for this resolution effect for the occupation fraction, we additionally calculate the occupation fraction for the higher resolution simulation with the condition, that one halo has to contain a stellar mass of at least $4\rm \,m_{target}$ (magenta) or $8\rm \,m_{target}$ (red line), which corresponds to the smallest/mean mass of one star particle in the lower resolution simulation.
    To produce this plot, we used the same method as for the production of \cref{fig:stellarFraction}.}
    \label{fig:stellarFractionZ4}
\end{figure*}

In all simulations presented in the main paper, we use a spatial zoom factor 8, equivalent to $8^3$ times better mass resolution than that of the parent box (see \cref{table:res}).
This enhanced mass resolution allows the smallest star-forming haloes with a mass of $10^6\msun$ to be resolved by almost 200 DM particles.
For the ``standard'' setup, which employs the radiation field from the \thesanone simulation as a boundary condition, we also conducted simulations with a zoom factor of 4. However, this lower resolution results in eight times worse mass resolution, meaning we can no longer resolve these small haloes. Consequently, this leads to a shift in star formation towards larger haloes. 
We first show the stellar-to-halo mass relation for both resolution levels in \cref{fig:stellarFractionZ4}. 
We used the same procedure as for the production of \cref{fig:stellarFraction}.
For halo masses above $10^8\msun$, both simulations yield similar results across all redshifts, with a slight difference observed at $z=5$.
For lower resolution, the occupation fraction shifts to higher halo masses.
Our definition of the occupation fraction itself depends on the mass resolution of the stellar component. 
To analyze the impact of this dependency, we additionally calculate the occupation fraction for the high-resolution simulation using the condition that one halo contains at least a stellar component of mass $\rm 4 m_{target}$ or $\rm 8 m_{target}$.
The first one is motivated by the minimum mass of a star particle of $\rm 0.5 m_{target}$ while the second condition uses the mean particle mass.
The first condition seems to reproduce the results of the lower resolution well for  $z \leq 5$, while the second condition works better for $z \geq 7$.

\begin{figure*}
    \centering
    \includegraphics[width=1\linewidth]{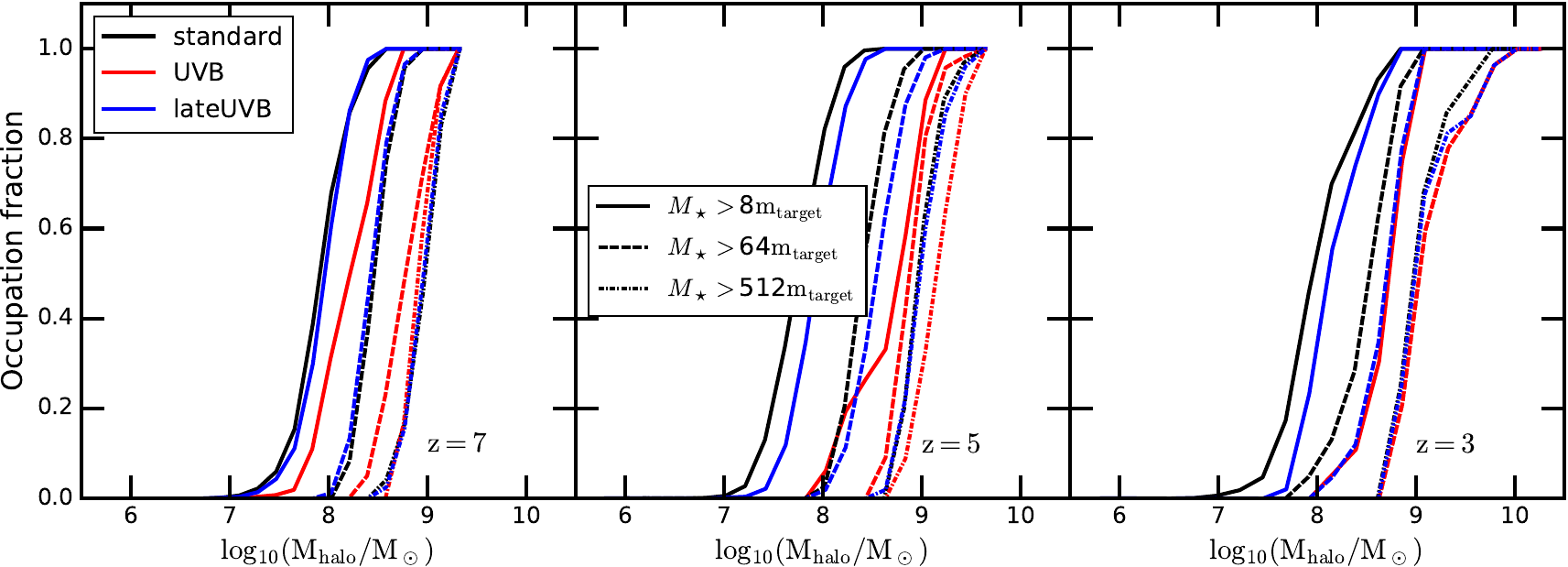}
    \caption{The occupation fraction as a function of halo mass at three different redshifts is analyzed for all simulations with a zoom factor of 8 and across three different simulation sets. 
    We employ three different definitions of the occupation fraction to emulate lower-resolution simulations based on the stellar mass of each halo: A stellar mass larger than 8/64/512 times the mean stellar particle mass $\rm m_{target}$ is required.
    As expected, the distribution shifts to larger halo masses as the threshold values increase. For $512 \, \rm m_{target}$ all three simulation sets show a similar occupation fraction.
    This corresponds approximately to the resolution of the original \thesanone parent box.}
    \label{fig:occupation_fraction_resolution_dependence}
\end{figure*}

We expect the three different simulation sets to show a similar distribution of the occupation fraction for low enough resolution since only high mass haloes can be resolved.
To analyze this effect, we show in \cref{fig:occupation_fraction_resolution_dependence} the occupation fraction as a function of halo mass for different threshold values for the stellar mass of a halo.
For $\rm M_\star > 512\, m_{target}$ the occupation fractions for all three simulation sets are almost the same.
This resolution corresponds to one star particle in the \thesanone simulation and highlights the importance of an increased resolution to analyze the impact of external UV radiation on low-mass haloes.

\begin{figure*}
    \centering
    \includegraphics[width=1\linewidth]{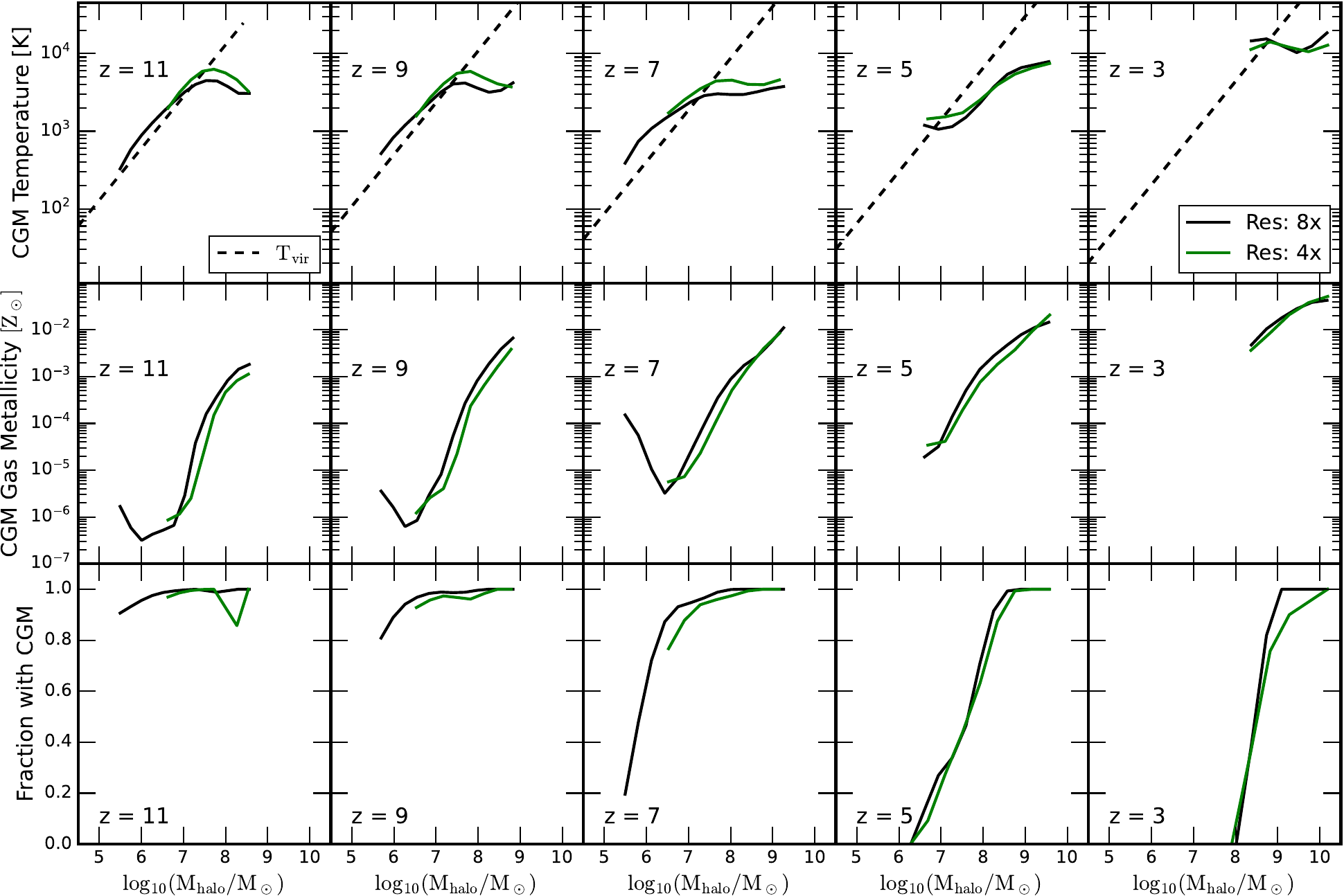}
    \caption{The impact of numerical resolution on the temporal evolution of the CGM as a function of halo mass.
    We show the results for the ``standard'' setup with a zoom factor 8 (black line, used in the main part of the paper) and a zoom factor 4 (green line).
    The lower zoom factor has eight times worse mass resolution. 
    We use the same method to define the CGM and calculate its properties as for \cref{fig:cgm_properties}.
    The simulations with lower resolution show slightly lower gas metallicity, particularly in haloes with masses below $10^9\msun$. This lower metallicity leads to less efficient cooling and slightly higher gas temperatures.}
    \label{fig:cgm_properties_z4}
\end{figure*}

To analyse the impact of resolution on the gas phase, we compare the properties of the CGM for both simulations in \cref{fig:cgm_properties_z4}.
We use the same definitions and procedures as for \cref{fig:cgm_properties}. 
The lower resolution simulation shows a slightly reduced gas metallicity in haloes below $10^9\msun$, leading to less efficient cooling and smaller deviations from the analytical virial temperature compared to the higher resolution simulation. 
The fraction of haloes with CGM remains very similar for both cases.

\section{The strength of the external radiation field in the different simulation sets}
\label{app:UV_strength}
\begin{figure*}
    \centering
    \includegraphics[width=1\linewidth]{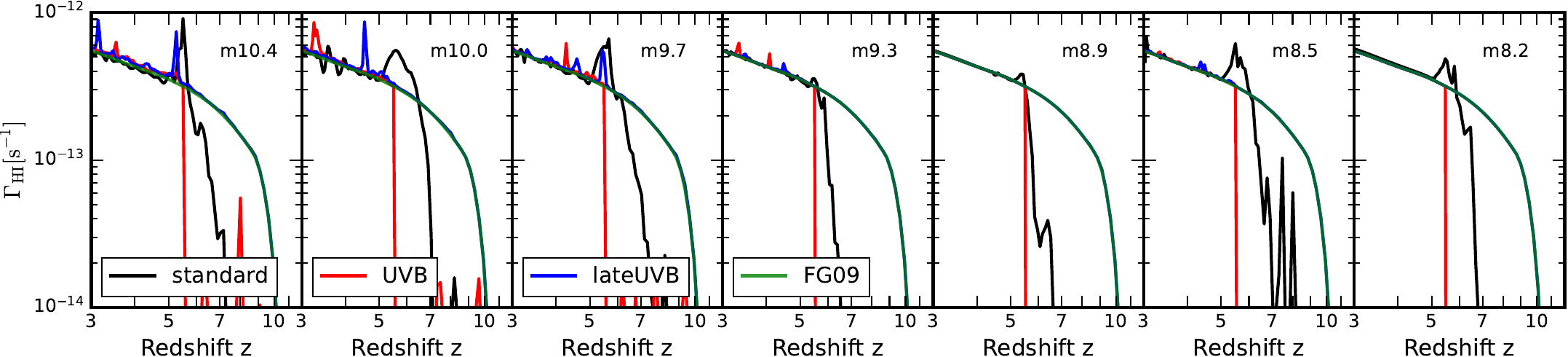}
    \caption{Median hydrogen photoionization rates in the high-resolution IGM for all seven zoom-in regions as a function of redshift.  
For the ``UVB'' and ``lateUVB'' sets, we add the uniform UV background from \protect\cite{FG09} to the local photoionization rate $\Gamma_{\mathrm{HI}}$ obtained from the radiation field, while in the ``standard'' set the external radiation field self-consistently propagates into the high-resolution region.  
For comparison, we also show the uniform UV background (``FG09''), which dominates the total photoionization rate in the ``UVB'' set across most zoom-in regions except during starbursts.
}
    \label{fig:uv_background}
\end{figure*}
In the ``standard'' simulation set, we use the UV radiation field extracted from the parent \thesanone\ box as a boundary condition up to redshift $z = 5.5$. By construction, this radiation field is spatially non-uniform, making a direct comparison to a uniform UV background more challenging.  
Nevertheless, to estimate its effective strength, we compute the median hydrogen photoionization rates in the high-resolution IGM, which we define as all gas not associated with any FOF halo.  
We include all three photon energy bins above $13.6\,\mathrm{eV}$, adopt a reduced speed of light of $0.01c$, and use the average ionization cross-sections for each bin as described in \cite{zoomIntro}.  
For the ``UVB'' and ``lateUVB'' simulation sets, we additionally include the uniform UV background from \cite{FG09} after its activation.  
The resulting hydrogen photoionization rates for all seven zoom-in regions are shown in \cref{fig:uv_background}.
The uniform UV background dominates the local radiation field in most zoom-in regions, except during starbursts in the largest ones. In the ``standard'' set, boundary photon transport is sufficiently efficient to establish a UV background throughout the full zoom-in region by $z = 5$.

\bsp	
\label{lastpage}
\end{document}